\documentclass{article}

\usepackage[verbose=true,letterpaper]{geometry}
\AtBeginDocument{
  \newgeometry{
    textheight=9in,
    textwidth=5.5in,
    top=1in,
    headheight=12pt,
    headsep=25pt,
    footskip=30pt
  }
}

\makeatletter
\newcommand*{\addFileDependency}[1]{
  \typeout{(#1)}
  \@addtofilelist{#1}
  \IfFileExists{#1}{}{\typeout{No file #1.}}
}
\makeatother

\usepackage{bm}
\usepackage{amssymb}
\usepackage[mathscr]{euscript}
\usepackage{amsmath}
\DeclareMathOperator{\Tr}{Tr}
\newcommand{\norm}[1]{\left\lVert#1\right\rVert}

\usepackage{hyperref}
\usepackage[nameinlink,capitalise,noabbrev]{cleveref}
\usepackage{graphicx}

\usepackage[utf8]{inputenc} 
\usepackage[T1]{fontenc}    
\usepackage{hyperref}       
\usepackage{url}            
\usepackage{booktabs}       
\usepackage{amsfonts}       
\usepackage{nicefrac}       
\usepackage{microtype}      
\usepackage{xcolor}         
\usepackage{float}

\title{CHARTING AND NAVIGATING THE SPACE OF SOLUTIONS FOR RECURRENT NEURAL NETWORKS}

\author{%
  Elia Turner \\
  Department of Mathematics \\
  Technion, Israel Institute of Technology\\
  \texttt{eliaturner@campus.technion.ac.il} \\
  Kabir Dabholkar \\
  Department of Mathematics \\
  Technion, Israel Institute of Technology\\
  \texttt{kabir@campus.technion.ac.il} \\
  Omri Barak \\
  Rappaport Faculty of Medicine and Network Biology Research Laboratory \\
  Technion, Israel Institute of Technology \\
  \texttt{omri.barak@gmail.com}
}

\begin{document}

\maketitle

\begin{abstract}

In recent years Recurrent Neural Networks (RNNs) were successfully used to model the way neural activity drives task-related behavior in animals, operating under the implicit assumption that the obtained solutions are universal. Observations in both neuroscience and machine learning challenge this assumption. Animals can approach a given task with a variety of strategies, and training machine learning algorithms introduces the phenomenon of underspecification. These observations imply that every task is associated with a space of solutions. To date, the structure of this space is not understood, limiting the approach of comparing RNNs with neural data.
Here, we characterize the space of solutions associated with various tasks. We first study a simple two-neuron network on a task that leads to multiple solutions. We trace the nature of the final solution back to the network’s initial connectivity and identify discrete dynamical regimes that underlie this diversity. We then examine three neuroscience-inspired tasks: Delayed discrimination, Interval discrimination, and Time reproduction. For each task, we find a rich set of solutions. One layer of variability can be found directly in the neural activity of the networks. An additional layer is uncovered by testing the trained networks' ability to extrapolate, as a perturbation to a system often reveals hidden structure. Furthermore, we relate extrapolation patterns to specific dynamical objects and effective algorithms found by the networks. We introduce a tool to derive the reduced dynamics of networks by generating a compact directed graph describing the essence of the dynamics with regards to behavioral inputs and outputs. Using this representation, we can partition the solutions to each task into a handful of types and show that neural features can partially predict them.
Taken together, our results shed light on the concept of the space of solutions and its uses both in Machine learning and in Neuroscience.
\end{abstract}

\section{Introduction}

Modern machine learning operates in an over-parameterized regime, implying that many different parameter-sets can achieve low error on a given training set  \cite{damourUnderspecificationPresentsChallenges2020}. This observation implies that for every task, there exists a space of solutions that can implement it. What are the properties of such a solution space? Are networks able to learn solutions that capture the intended underlying phenomena or do they reach artificial shortcuts that do not generalize well?
What biases networks to prefer one solution over the other? These questions remain largely unanswered.
A parallel phenomenon occurs in Neuroscience. When animals are instructed to perform a task in a controlled environment, they exhibit both neural and behavioral variability, which stem from different task-strategies \cite{mcintyrePatternsBrainAcetylcholine2003,gholamrezaeiIndividualDifferencesSkilled2009,patitucciOriginsIndividualDifferences2016, kempermannEnvironmentalEnrichmentNew2019, musallHarnessingBehavioralDiversity2019, kurikawaNeuronalStabilityMedial2018, moritaPsychometricCurveBehavioral2011, castroHumansDeployDiverse2013, lebovichIdiosyncraticChoiceBias2019, giladBehavioralStrategyDetermines2018}.
In Computational Neuroscience trained Recurrent Neural Networks (RNNs) are used as a tool to explain functions and mechanisms that are observed in brain dynamics \cite{barakRecurrentNeuralNetworks2017}. In fact, various recent studies have matched the activity of trained RNNs to that of experimental recording \cite{barakFixedPointsChaos2013, manteContextdependentComputationRecurrent2013,sussilloNeuralNetworkThat2015,yangTaskRepresentationsNeural2019, schrimpfNeuralArchitectureLanguage2020, vyasComputationNeuralPopulation2020, remingtonFlexibleSensorimotorComputations2018}.
In light of the variability that undoubtedly exists on both sides of the comparison, these results seem puzzling. In this work, we present multiple tasks for which trained RNNs produce a rich space of qualitatively different solutions. We argue that to properly use artificial networks, and RNNs in particular, as models of neural circuits that support a given task, it is necessary to chart the space of solutions that arises from training.

Here, we first apply this approach to a simple two-neuron network and demonstrate how distinct solutions arise. We then study three tasks inspired by the neuroscience literature: interval reproduction  \cite{jazayeriTemporalContextCalibrates2010}, delayed discrimination \cite{romoNeuronalCorrelatesParametric1999a}, and interval discrimination \cite{biUnderstandingComputationTime2020}. We show that different networks with identical hyperparameters find qualitatively different solutions. We find one layer of variability within the neural activity in response to stimuli from the training set. Since by design the output of all networks is identical during training, this layer is akin to multiple realizability \cite{guestLogicalInferenceBrains2021}. Next, we expose an additional layer of variability when we challenge the networks with inputs that are outside the distribution of the training set and systematically characterize the responses. Furthermore, we manage to show that the diversity revealed with these challenging inputs corresponds to qualitatively different computations performed by the network. To chart the space of solutions, we introduce a tool that reduces the dynamics of a network into a graph that captures the essence of the computation performed. Applying it to all networks partitions the space into a handful of possible reduced dynamics. Additionally, these classes can be partially predicted using experimentally accessible neural activity obtained only in response to trained stimuli.

\section{Related work}
Recent work, \cite{NEURIPS2019_5f5d4720} 
studied the effects of modeling choices over the dynamics of the trained solutions, in a few canonical tasks. They trained thousands of RNNs, while systematically controlling the hyperparameters, and analyzed the geometrical and topological aspects of each solution. They concluded that while the geometry of different solutions can vary significantly across different architectures, the underlying computation and dynamical objects are widely universal. Our results are superficially opposite to theirs. This is probably due both to the choice of tasks and to challenging networks with unexpected inputs.

The authors of \cite{mehrerIndividualDifferencesDeep2020} show that deep networks trained on vision tasks develop different internal representations, especially in the higher layers. While reaching similar conclusions on the need to use populations of networks for neuroscience comparisons, the analysis of feedforward networks naturally focuses on representations rather than on dynamics.
While writing this manuscript, we were made aware of recent work by \cite{ghazizadehSlowManifoldsNetwork2021} that takes a very similar approach to ours. Apart from the different tasks, architectures and learning rules studied there, our analysis of the two-neuron network also provides a tangible example of discrete basins of attraction in solution space.
An underspecified two-dimensional epidemiological model was studied by \cite{damourUnderspecificationPresentsChallenges2020}. The variability in solutions there, however, was continuous. Specifically, all solutions were connected on a single manifold and the dynamics of the system did not undergo bifurcations along this manifold.

Underspecification was also highlighted in neuroscience models that are not based on artificial neural networks \cite{marderMultipleModelsCapture2011} and is observed more generally in  complex systems \cite{machtaParameterSpaceCompression2013}.

\section{What constitutes a solution?}
Before comparing how the different networks solve the various tasks, it is worth dwelling on the concept of a solution. This is not a trivial concept and can be related to fundamental philosophical questions. Aristotle suggested \cite{falconandreaAristotleCausality2019} the four causes of understanding an object: its material form, its formal description, its efficient cause (creation process), and its final cause (purpose). We can draw an analogy to our understanding of RNN solutions, and ask: What is their architecture (material), what is their underlying algorithm or dynamics (formal), which optimization process led to their final state (efficient), and what task do they solve (final).
Because our motivation stems from comparing networks to biological data, we take an operative approach that relies on measures that could in principle be obtained experimentally. We thus consider two of these pillars: either the neural activity of the network while performing the task (formal description), or its predictions for unexpected stimuli (purpose).

\section{Space of solutions in a 2D RNN task}
\begin{figure}[h]
\begin{center}
\includegraphics[width=1\textwidth]{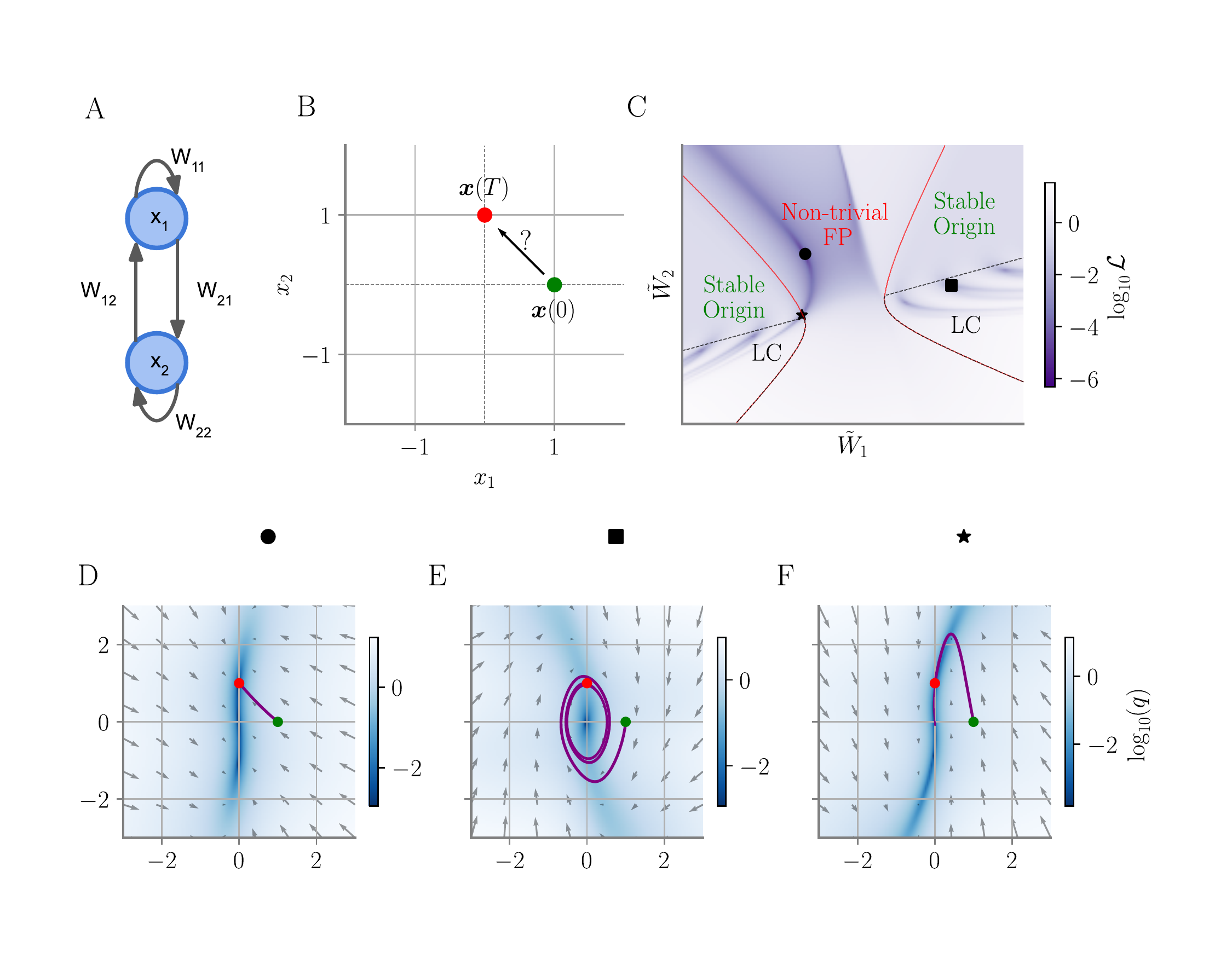}
\end{center}
\caption{Two-neuron network. \textbf{A} Network architecture, with the four trainable parameters. \textbf{B} The task defined in phase space. The initial state is fixed (green). The task is for the network to be at a final state (red) at time $T=10$. \textbf{C} A 2-D slice of the 4-D parameter space containing three selected solutions (black markers). Heatmap denotes task loss evaluated along the slice. Overlaid is a bifurcation diagram along the slice; lines indicate dynamical bifurcations and text indicates regions with a stable origin, non-trivial fixed point attractors and limit cycles (LC). \textbf{D, E, F} Phase portraits of the three selected solutions marked in \textbf{C}, trajectories taken by each of the networks during the task ($t<T$, thick purple) and subsequently ($t>T$, thin purple). Heatmaps denote speed of dynamics $q:=\frac{1}{2}\norm{\boldsymbol{\dot{x}}}^2$\cite{sussilloOpeningBlackBox2013}}.
\label{fig:2D}
\end{figure}

We first study the space of solutions in an extremely simple scenario that nevertheless shows qualitatively different solutions. We consider a 2-neuron continuous-time network (\cref{fig:2D}A), in which the state of the network $\boldsymbol{x}\in \mathbb{R}^2$ evolves according to:

\begin{align}
    \boldsymbol{\dot{x}}=-\boldsymbol{x}+W\phi(\boldsymbol{x}), \
    \boldsymbol{x}(0) = \begin{bmatrix}1 \\ 0 \end{bmatrix}
\end{align}
 
 For the task,  we require that $\boldsymbol{x}(T) =\begin{bmatrix}0 \\ 1 \end{bmatrix}$ (\cref{fig:2D}B). We choose $\phi:=\tanh$ and $T=10$. Accordingly we define a loss:
 
 \begin{align}
	\mathcal{L} = \frac{1}{2} \norm{\boldsymbol{x}(T)- \begin{bmatrix}0 \\ 1 \end{bmatrix}}^2.
\end{align}

The system has four parameters, given by the elements of $W\in \mathbb{R}^{2\times2}$. We train 10000 randomly initialized networks to minimize $\mathcal L$ and discover solutions with qualitatively different dynamics. We select three representative examples. (\cref{fig:2D} D,E,F). The first example implements a stable fixed point near the target, the second implements a limit cycle near the target, and the third exhibits transient amplification passing through the target before gradually decaying to zero.

To visualize the space of solutions, we consider a 2D plane in the 4D parameter space containing the three aforementioned solutions. We find that this plane explains 94 percent of the variance of all solutions (compared to 95 percent by the first two principal components), and is thus a representative description of solution space. The shading in Figure 1C shows the loss along this plane, indicating that some solutions lie along a single continuous manifold, while others inhabit discrete islands. Linearizing the dynamics at the origin allows us to obtain a bifurcation diagram on this plane (Figure 1C, red and black lines), showing the existence of a bifurcation along the continuous manifold. In the simpler case of a linear network, we can also show how the discrete and continuous solution sets arise (see supplementary section 2.1)

We thus conclude that the network converges to qualitatively different solutions. These can be attributed to discrete basins of attraction in parameter space and dynamical bifurcations of the system occurring within and across these basins. 

\section{Neuroscience Tasks}
\begin{figure}[h]
\begin{center}
\includegraphics[width=1\textwidth]{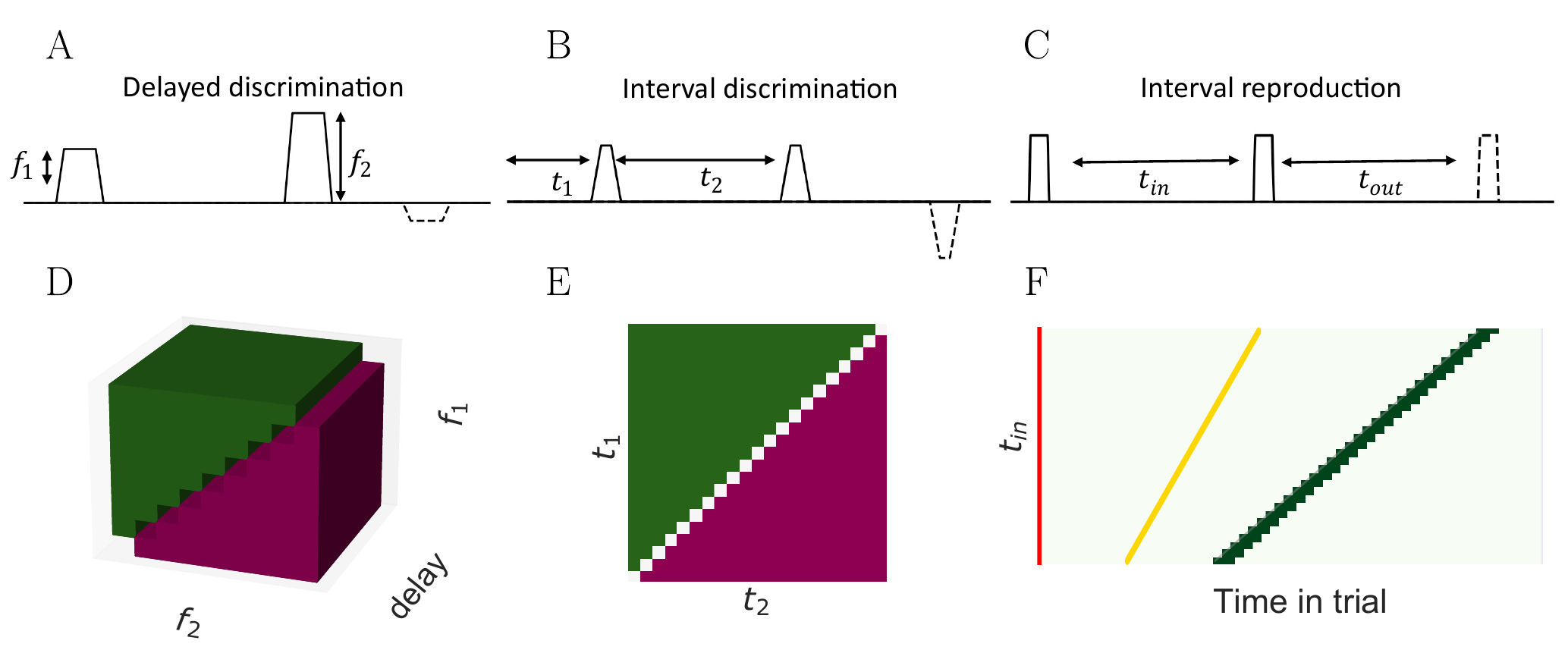}
\end{center}
\caption[Tasks]{The three tasks. \textbf{A} Delayed discrimination task. Top: The network is presented with two pulses with amplitudes $f_1$ and $f_2$ respectively, separated by a variable delay, and should respond with a pulse with an amplitude equal to $\text{sign}(f_1-f_2)$. Bottom: The desired output for all training data, for each delay, $f_1$, $f_2$ the color of the 3D matrix is the expected response. \textbf{B} Interval discrimination task. Top: The network is presented with two pulses with a unit amplitude, that arrive at times $t_1$ and $t_2-t_1$. In response, it should produce a pulse with amplitude that is equal to $\text{sign}(t_2-t_1)$. Bottom: Desired output for all $t_1$ and $t_2$ combinations. \textbf{C} Interval reproduction (Ready-Set-Go) task. Top: The network is presented with two pulses (Ready and Set, provided through separate input channels) separated by $t_{in}$ time-steps. The desired output is a Go pulse, delayed by $t_{out} = t_{in}$ steps from the Set pulse. Bottom: Desired output for all trials, with Ready, Set and Go depicted in red, yellow and green respectively..
}
\label{fig:task}
\end{figure}

Guided by the results from the two-neuron network, we examine three more complex, neuroscience-inspired, tasks \cref{fig:task}. A priori, it is not clear whether such tasks will exhibit more or less variable solutions. On the one hand, a complex task might lend itself to multiple algorithmic solutions. On the other hand, a complex task represents more constraints on the network and hence might lead to convergence to a unique solution. More details can be found in the supplementary material subsection \cref{appendix:taskdata}

In the \textbf{Delayed discrimination} task \cite{romoNeuronalCorrelatesParametric1999a}, two pulses of varying amplitudes ($f_1,f_2 \in [2,10]$) are presented, separated by a varying delay ($t_d \in [5,24]$) \cref{fig:task}A. The lower panel shows the correct output in response to various stimulus combinations, which is independent of $t_d$ and partitions the $(f_1,f_2)$ plane. 

In the \textbf{Interval discrimination} task \cite{biUnderstandingComputationTime2020}, two pulses of equal amplitude are presented at times $t_1$ and $t_1 + t_2$, where $t_1, t_2\in [10,30]$ and $t_1 \neq t_2$. The network should produce an output pulse whos sign indicates whether $t_2>t_1$ or not \cref{fig:task}B. The lower panel once more shows the desired output.

Finally, in the \textbf{Interval reproduction} task \cite{jazayeriTemporalContextCalibrates2010} the network receives two input pulses -- Ready and Set -- separated by $t_{in}$ time steps. The task is to generate an output pulse $t_{out}=t_{in}$ time steps after the Set pulse. The training intervals were drawn from a uniform distribution $t_{in}\in [20, 50]$. The lower panel shows the desired output, where having only one parameter defining the trials ($t_{in}$) allows us to display the entire trial output on each line. Trials aligned to the \emph{Ready} pulse (\cref{fig:task}B, red). This results in the \emph{Set} pulse (yellow)  forming a line with slope $1.0$ and the \emph{Go} pulse (green) a line with slope $0.5$. \Cref{fig:task}C shows the output of a trained network matching this pattern.

For each task we trained $400$ Vanilla networks, with $N=\{20,30,40,50\}$ hidden units. See the Supplementary material subsection\cref{appendix:training} for more details regarding the training process, as well as results from GRU and LSTM networks trained on the interval reproduction task.

\begin{figure}[H]
\begin{center}
\includegraphics[width=0.8\textwidth]{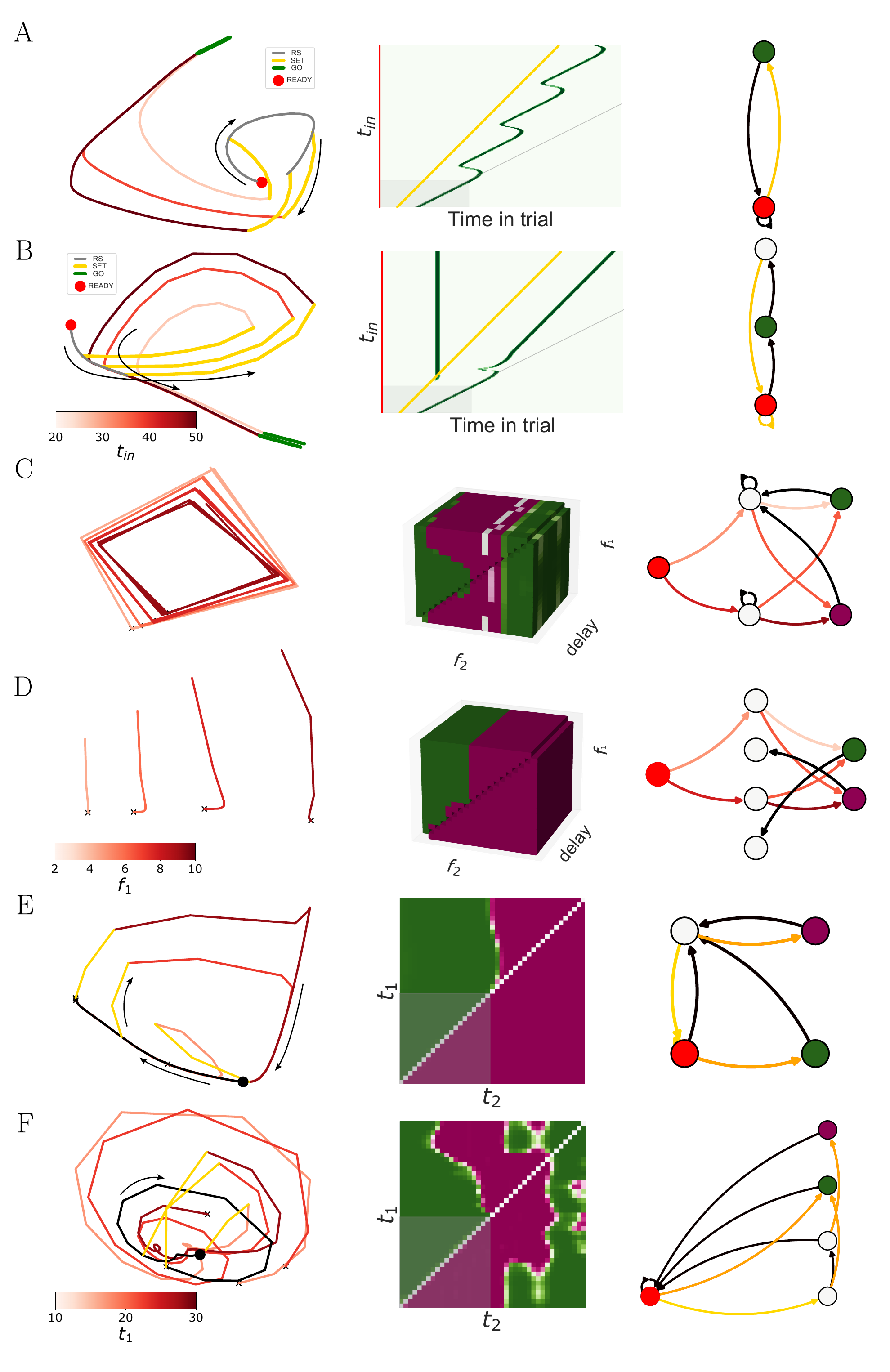}
\end{center}
 \caption[Task dynamics]{Diverse solutions for the same tasks. Two example networks are shown for each task (rows). Left column: two-dimensional PCA of the network activity. Middle column: Network output in response to extrapolation across task parameters. Right column: the reduced dynamics of the networks. \textbf{(A,B)} Solutions to the time production task. The PCA plots show the activity from the Ready pulse (red dot and black line), as well as from the Set pulse (yellow) up to the beginning of the output production, for three different task parameters (colorbar). Note that in A the Ready-Set epoch is separated from the Set-Go, whereas in B these epochs converge. \textbf{(C,D)} Solutions to the delayed discrimination task. The PCA plots show the activity between the two input pulses, for various $f_1$ amplitudes. The memory of the different amplitudes is kept either via limit cycles (C) or slow points (D).
 \textbf{(E,F)} Solutions to the interval discrimination task. The PCA plots show activity from trial onset (black dot and line), as well as from the first pulse (yellow) until the maximal delay, for three different $t_1$ values (colorbar). Note the convergence of the two epochs to the same trajectory in E, similar to B. }
 \label{fig:examples}
\end{figure}

\subsection{Multiple neural dynamics solve the same task}

We demonstrate the diversity of neural dynamics in different solutions by showing a couple of examples from each task. The left side of \Cref{fig:examples}AB  shows the activity of two networks solving the interval reproduction task projected onto the first principal components. It is possible to infer two distinct algorithms for solving the timing problem from these plots. Because the task requires counting time in two distinct epochs -- from Ready to Set and from Set to Go -- we focus on these to explain the algorithms. The network in panel B uses the same phase space trajectory for both epochs, effectively counting from the maximal delay ($t_{in}=50$) downwards upon the Ready pulse, and moving to an earlier point upon the Set pulse. This is in contrast to the network in panel A, in which the Set pulse leads to a different area in phase space. Furthermore, the almost circular trajectory of the Ready-Set epoch hints at the oscillatory behavior that will be discussed below.

The delayed discrimination task also admits multiple solutions. In this case, we consider activity during the delay period. \cref{fig:examples}CD show several delay trajectories, corresponding to different $f_1$ values. The network in panel D converges to approximate fixed points for each such value, while the network in panel C converges to limit cycles. Note that the network was trained with variable delays \cite{orhanDiverseRangeFactors2019a}, and hence the second stimulus arrives at random phases of these cycles. Nevertheless, both networks perform well on the training set.

Interval discrimination contains two timing-epochs -- from the onset of the trial to the first pulse, and between the two pulses. The network in panel E uses the same trajectory in phase space for both epochs, similar to panel A of the interval reproduction task. The second example exhibits both oscillatory behavior and distinct trajectories for the two phases. 

Taken together, these examples show that networks trained with identical hyperparameters, and reaching similar performance nevertheless develop qualitatively different neural dynamics to solve the same task.

\subsection{Extrapolation}
On the behavioral side, we challenge networks with stimuli that are outside of the trained distribution. Since each of our tasks is parametric, testing the networks on extrapolation while increasing each parameter is a natural choice. The middle column of \cref{fig:examples} shows the results of this challenge for the various networks, which often shows traces of the neural diversity described above.

The almost circular Ready-Set neural trajectory of panel A apparently results in a limit cycle, as revealed by the extrapolation plot. Similarly, the fact that the Ready-Set trajectory of panel B leads directly to the Go pulse is reflected in the vertical Go line that precedes the Set pulse in the extrapolation plot. Other features, such as the fact that a Set pulse delivered \textit{after} the Go pulse results in a second output, are only revealed through extrapolation and cannot be deduced from the PCA of neural activity during the training set.

In the delayed discrimination task, the oscillations shown in panel C were not reflected in the output on the training region. Extrapolating to larger amplitudes and delays, however, reveals output oscillations as depicted on the top face of the extrapolation cube.

In the interval discrimination task (panels E,F), the relation between neural trajectories and extrapolation behavior is less clear. But in this case, as in all others, the striking differences between different solutions are manifested both in the training neural activity and in the response to behavioral challenges.

\subsection{Reduced Dynamics}
Part of the motivation for looking into extrapolation patterns is uncovering the algorithm that networks use to solve the task. Within the framework of neural dynamics \cite{vyasComputationNeuralPopulation2020, barakRecurrentNeuralNetworks2017} this corresponds to mapping the dynamical objects used by the network and their relation to behavioral inputs and outputs. Previous works focused on fixed point topology to achieve this goal \cite{sussilloOpeningBlackBox2013, NEURIPS2019_5f5d4720, manteContextdependentComputationRecurrent2013}. The tasks considered in the present work are mostly dependent on transient dynamics and hence require a different approach. To this end, we developed a tool to compress the state space while preserving the essential dynamics and the behavioral information. Briefly, the method generates trajectories by combining long stretches of autonomous dynamics with behavioral inputs. The trajectories are then merged and compressed into a graph. Each node represents areas in phase space that have a given output and share the same input-dependent past and future. The edges represent autonomous dynamics (black) or the various inputs (colors). The full details of the algorithm are in the supplementary material subsection $2.2$. The left column of \cref{fig:examples} shows such reduced dynamics for all the example networks. Examining these graphs can reveal the dynamics driving extrapolation patterns. 

The network of panel A has a limit cycle, which is depicted by the red node that has a black self-loop. Furthermore, the graph shows that after the Go pulse, the network can still respond to an additional Set pulse. The fact that the network of panel B utilizes the same trajectory for both epochs is reflected in the yellow self-loop in the corresponding graph. The graph also shows that a Go pulse can occur before the Set pulse.

The delayed discrimination task has more input types and therefore was less compressed by the algorithm. As a result, the graphs can show both the logic of comparison during the training set and the extrapolation patterns with the limit cycles. A Similar scenario holds for the interval discrimination task.

\subsection{A space of solutions}
\begin{figure}[h]
\begin{center}
\includegraphics[width=1\textwidth]{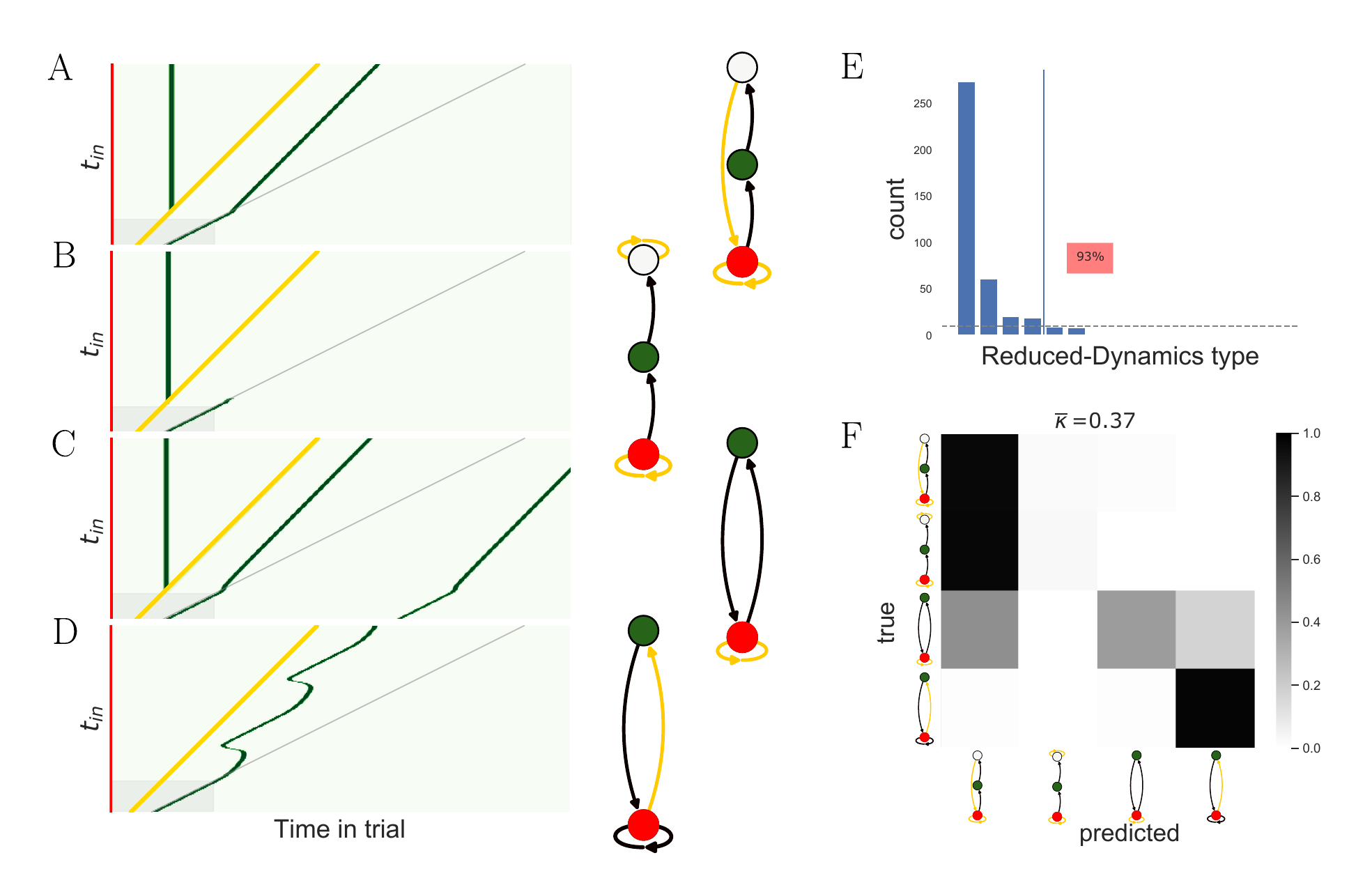}
\end{center}
 \caption[Space of solutions]{The space of solutions for the interval reproduction task.  \textbf{A-D} Representative extrapolation plots (left) and reduced dynamics graphs (right) for the four most common solution types. \textbf{E} Distribution of solution types for the $400$ networks trained. The four solutions shown account for $93\%$ of the networks. $F$ Neural features obtained during the training set can partially predict the solution type that includes extrapolation dynamics. The confusion matrix shows the result of this prediction. Note that the first two solution types are mixed by this prediction, but their dynamics during the training intervals is similar and they only differ in the dynamics after the Go pulse. }
 \label{fig:confusionmatrixrsg}
\end{figure}

The examples described above suggest that similar to the two-neuron case, trained networks converge into distinct solutions which can be characterized both by neural signatures and by behavioral ones. In the case of the two-neuron network, the parameter space is only four-dimensional, and most of the solutions are in a two-dimensional subspace. Hence, we could span most of it and understand the solution types via the linearized dynamics around the origin. In the more complex tasks, the reduced dynamics graphs serve as a method to characterize the space of solutions. We computed the reduced dynamics for all 400 networks of each task and found that they only contain a few different graphs. \cref{fig:confusionmatrixrsg}E shows the number of networks of each type for the interval reproduction task. The four major graphs and their corresponding representative extrapolation plots are shown in panels A-D.  

\subsection{Inferring reduced dynamics from neural activity}

The reduced dynamics described above are obtained by simulating the networks for very long times, to obtain their asymptotic behavior. In a neuroscience setting, such information is not readily available. Is it possible to infer this asymptotic behavior from neural activity during the training set? To answer this question, we extracted a set of features, as detailed in the supplementary material. The features mostly measure the relationship between neural activity in the different epochs of the task. Other features measure geometrical properties of a single epoch, as in the curvature of the Ready-Set trajectory, which was inspired by networks like \cref{fig:examples}A. While we tried to include features that are natural descriptors of the neural activity, there is some arbitrariness in our choice. We chose to err on the side of including more features, and later rely on cross-validation to avoid overfitting in our predictions \cite{khanPredictingOdorPleasantness2007}. 

For each task, we trained and evaluated a classifier to predict the reduced dynamics of the pool of networks from the neural features. \cref{fig:confusionmatrixrsg}F shows the resulting confusion matrix for the interval reproduction task (other tasks are in the supplementary material \cref{fig:romoconfusion,fig:tromoconfusion}). Cohen's kappa \cite{mchughInterraterReliabilityKappa2012} was used to summarize the prediction quality, showing an above-chance performance.

\subsection{Different architecture and tasks}

\begin{figure}[h]
\begin{center}
\includegraphics[width=1\textwidth]{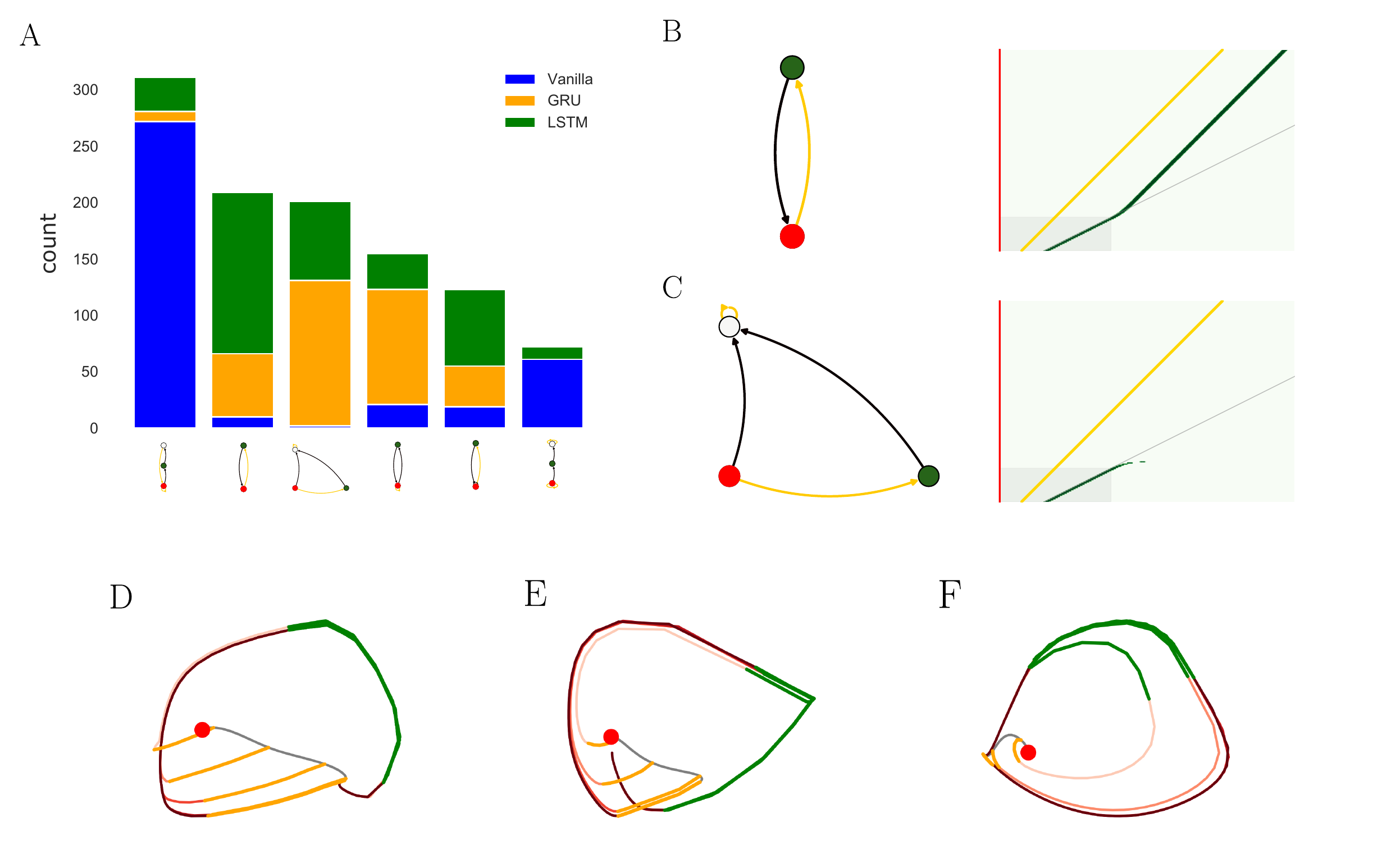}
\end{center}
\caption{Architecture biases, but does not determine the solution. \textbf{A} A histogram of the six most common reduced dynamics across all three architectures for the time-reproduction task, shown by stacking the architecture-specific histograms on top of one another.
\textbf{B, C} The reduced dynamics (left) and the extrapolation patterns (right) of the second and third most common solutions across all architectures, but rarely occur within the Vanilla networks.
\textbf{D, E, F} 
The 2D PCA of the dynamics of three networks from all three architectures (Vanilla, GRU, LSTM), that reach the solution shown in panel \textbf{B}. }
\label{fig:bias}
\end{figure}

The diversity of solutions reported here arise from identical hyperparameters and random initialization. To probe the biases introduced by changing hyperparameters, we tested the effect of varying network architecture for the interval reproduction task. The supplementary material shows results from training this task using LSTM and GRU networks. We find that some solution classes are shared between architectures, and some only appear in certain architectures (\cref{fig:bias}A). Furthermore, these choices bias the solutions, as reflected in the different histograms. For instance, the example of \cref{fig:bias}B did not occur in any of the $400$ Vanilla networks, but was the most common type in GRU networks. Despite these biases, there are solutions that appear under all architectures, as exemplified in \cref{fig:bias}C,D,E.


The chosen tasks so far emphasized variability. We also examined a context-dependent integration task, which was previously shown to exhibit a universal solution \cite{NEURIPS2019_5f5d4720}. The supplementary material shows an analysis of the response of networks to various behavioral challenges. We find that there is considerable variability in the response of networks to challenges, but of a quantitative rather than a qualitative nature. Furthermore, we also quantified various aspects of neural variability in these networks. We did not find significant correlations between the various behavioral and neural measures, suggesting that there are many axes of individuality for these networks.

\section{Discussion}
In recent years, trained RNNs were successfully used to model biological circuits. Specifically, these networks converged to solutions that were similar to those of their biological counterparts. This observation is puzzling for any complex system and particularly regarding the brain, a complex biological system in which variability is the rule rather than the exception. 
Inspired by this puzzle, we study the space of solutions of RNNs for both a simple two-neuron network and three neuroscience-inspired tasks.
Through an interplay between observation, analysis, and perturbation in the form of extrapolation, we discovered for each task qualitatively different solutions. By analyzing the neural activity of these networks, we observed that the variety of behavioral phenotypes originate from only a handful of distinct dynamical mechanisms.
To affirm these observations and describe the essence of each solution, we introduced a tool that summarizes the network computation in a reduced-dynamics graph that contains the essence of the dynamics as they relate to behavioral inputs and outputs.
By extracting neural features from the training activity, we showed that the solution type can be partially predicted from experimentally accessible measurements.

We show using a 2D continuous-time RNN that qualitatively different dynamical topologies can arise in the context of a single task and that learning can find different solutions depending on initialization. In a related study, the authors analyze a low-D continuous-time GRU and find a large diversity of dynamical topologies \cite{jordanGatedRecurrentUnits2021}. They find that the 2D GRU finds topologically different solutions to variants of a working memory task depending on the input statistics. Such a modification to the task, although subtle, modifies the loss landscape and learning dynamics. In our work we emphasize that such topological diversity of solutions can arise for the same task, that is to say within the same loss landscape. Both our work and theirs identify the implications of topological diversity for generalization.

Asking a seemingly trivial question - "what is a solution?", highlights a complex and vital topic.
Even though artificial and biological networks are being compared regularly, and there is a common intuitive understanding of what properties are relevant for this comparison, the precise definitions of "solution" or "mechanism" are rarely discussed \cite{guestLogicalInferenceBrains2021}. Is dynamical similarity within the training regime sufficient for asserting two solutions are alike? What is the proper way to test subjects' performance on a learned task? Can we answer these questions while considering the specifics of each task? Under what conditions can we safely assume that there exists a universal solution? We believe that every study that models neural circuits should be explicit about such meta-concerns. In this work, we charted solutions with an operational approach in mind and considered solutions as different as long as they qualitatively differ either in their neural activity within the training set or in their behavioral prediction on extrapolation trials. The reduced dynamics tool is our attempt to describe and differentiate solutions according to the criteria that we consider important. 

Our work has several limitations that should be noted. We focused on the variability arising with identical hyperparameters and therefore did not systematically explore the effect of changing them. Specifically, how network size, learning algorithm, and choice of architecture bias the solutions remains to be explored.

The space of solutions for a given task is, above all else, a property of the task itself. We showed examples of three tasks that have substantial variability, and of one task (context dependent integration) that has much less variability. Yet, we do not know why some tasks admit multiple solutions and others do not. We conjecture that the tasks presented in \cite{NEURIPS2019_5f5d4720} require an informative output at all points in time, whereas the tasks we considered require an informative output only once at the end of the trial.

 The reduced-dynamics tool relies on a set of trajectories as a starting point. We opted to include  regions in phase space that are reachable with task-consistent inputs (similar to \cite{sussilloOpeningBlackBox2013}). While exploring the full phase space is not feasible, our choice may limit the description obtained by our tool. 
 Finally, the space of solutions was described for networks that have already learned to perform a task. It remains to be seen whether it can be used as a map in which to understand how the process of learning itself takes place.

To conclude, we found that RNNs can produce a diverse set of solutions to the same computational tasks. These solutions represent distinct algorithms and are supported by corresponding dynamical objects. The solutions are isolated in parameter space, causing the initial conditions to largely determine the outcome. Furthermore, we showed that experimentally accessible tools can be used to indirectly characterize the asymptotic properties of the solutions. We believe that exploring the space of solutions can advance neuroscience, machine learning, and their intersection --  making more rigorous comparisons of models to data.

\section{Acknowledgements}

This work was supported in part by the Israeli Science Foundation (grant number 346/16, OB) and by a Rappaport institute thematic grant (OB).


\bibliography{turner_charting}
\bibliographystyle{IEEEtran}

\newpage
\appendix{}

{ \Large Supplementary material }

A python code for the figures and results is partially available at \url{https://github.com/eliaturner/space-of-solutions-RNN/}.

\section{Two-neuron RNN}

\subsection{Discrete and continuous solution manifolds in the linear two-neuron RNN}
In this section, we analyze a simplified linear discrete-time version. This allows an analytical solution of the weights required to perform the task and demonstrates the nature of the space of solutions. The equations are now

\begin{align}
	\boldsymbol x_{t+1} &=  W \boldsymbol x_t \\
	\boldsymbol x_0 &= \begin{bmatrix}1 \\ 0\end{bmatrix}
\end{align}
where $W \in \mathbb R^{2 \times 2}$.

The task requires that $x_n=\begin{bmatrix}0 \\ 1\end{bmatrix}$. This can be solved either analytically (below) or numerically by minimizing the following loss by running gradient descent on $W$ from an initial $W_0$:
\begin{align}
	\mathcal L = \norm{x_n-\begin{bmatrix}0 \\ 1\end{bmatrix}}^2
\end{align}

We observe that different random initializations converge to different points in the space of solutions. In this document, we try to describe and visualize this space.

\subsubsection{The space of solutions}
We need solutions to the following equations:
\begin{align}
	\label{eq:system1}
	\begin{bmatrix}0 \\ 1\end{bmatrix} &= A \begin{bmatrix}1 \\ 0\end{bmatrix} \\
	\label{eq:system2}A &= W^n
\end{align}
This breaks down the problem into a feedforward problem of solving \eqref{eq:system1} for $A$ and a recurrent problem of solving \eqref{eq:system1} and \eqref{eq:system2} for W.

The space of solutions to the feedforward problem is simple: 
\begin{align}\mathcal A:= \{\text{$A \in \mathbb R^{2 \times 2}$ such that $A_{11}=0$ and $A_{21}=1$}\}.\end{align} 

To solve the recurrent problem, we need to additionally find the n\textsuperscript{th} real roots of $A \in \mathcal A$ if they exist. For simplicity, we focus on a task of only two time steps $n=2$.

\subsubsection{Solutions of a two time step task}
We begin by considering only nonsingular $A$ ($A_{12}\neq 0$ in this case).
The existence and number of real $W$ (branches), such that $A=W^2$ is given by spectrum of $A$. If $A$ has any real negative eigenvalues, then there are no real solutions. If $A$ has no real negative eigenvalues then $A$ has $2^{r+c}$ square roots where $r$ is number of real eigenvalues and $c$ is the number of complex eigenpairs (\cite{higham1987computing} Theorem 7).

First we study the case of $A$ with complex eigenvalues $\theta\pm i\mu$ with $\mu\neq 0$. 
\begin{align}
	\mathcal A_{\mathbb C} &= \{A \in \mathcal A \text{ such that Tr}(A)^2-4\det(A)<0\} \\
			       &= \{A \in \mathcal A \text{ and } A_{22}^2+4A_{12}<0 \}
\end{align}

$A \in \mathcal A_{\mathbb C}$ has $2^1$ real roots given by $W = \pm(\alpha \boldsymbol 1 + \frac{1}{2\alpha}(A-\theta \boldsymbol 1))$ where $(\alpha+i\beta)^2=\theta + i\mu$. (Eq 4.6 in \cite{higham1987computing})

Next we identify the the $A$s with eigendecomposition $A=V\begin{bmatrix} \theta_1 & 0 \\ 0 & \theta_2\end{bmatrix}V^{-1}$ where the $\theta_i$s are real and positive.
\begin{align}
	\mathcal A_{\mathbb R,>0} &= \{A \in \mathcal A \text{ such that }\Tr(A)^2-4\det(A)>0 \text{ and } \Tr(A) > \sqrt{\Tr(A)^2-4\det(A)}\} \\
				  &= \{A \in \mathcal A \text{ and } A_{22}^2+4A_{12}>0 \text{ and } A_{22}>0 \text{ and } A_{12}<0 \}
\end{align}

There are $2^2$ real roots of $A \in \mathcal A_{\mathbb R, >0}$ given by $W=V\begin{bmatrix}\pm \sqrt{\theta_1} & 0 \\ 0 & \pm\sqrt{\theta_2}\end{bmatrix}V^{-1}$ taking each of the four combinations of the $\pm$s.

\begin{figure}[ht!]
\centering
\includegraphics[width=.95\linewidth]{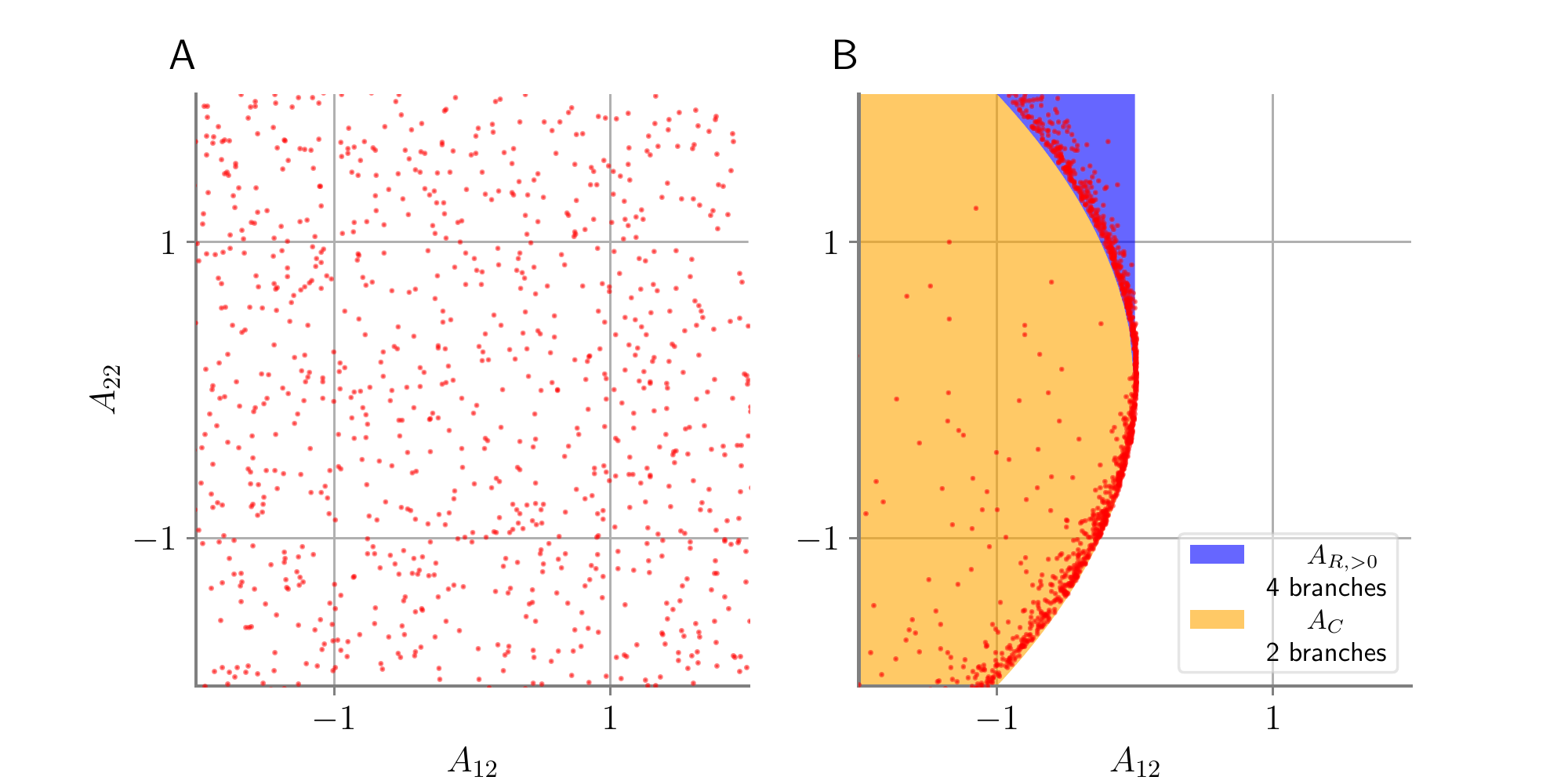}
\caption{\label{fig: feedforward recurrent}Solutions of gradient descent from random initialization to the feedforward problem (A) and the recurrent problem (B) in the space of $A$ matrices. We only plot the two unconstrained elements $A_{12}$ and $A_{22}$.}
\end{figure}

We numerically verify these statements by running gradient descent from random initialization $W_0^{(i,j)}\sim U(-3,3)$. Figure \ref{fig: feedforward recurrent} shows these solutions on the non-zero elements of the $A$ matrix. The left panel shows the continuous nature of the feed-forward problem. In contrast, the right panel shows a highly non-uniform distribution, matching the regions predicted from the analysis above. 

We see that while the feedforward problem has a plane of degeneracy, the recurrent problem has two modes of degeneracy: firstly solutions fall somewhere in a subset of the same feedforward plane of degeneracy and secondly, at each point on the plane the solution can fall on one of several branches. While the first mode is continuous in nature, the second mode is disjoint.

\subsubsection{Visualizing the Recurrent solutions}
\begin{figure}[ht!]
\centering
\includegraphics[width=.95\linewidth]{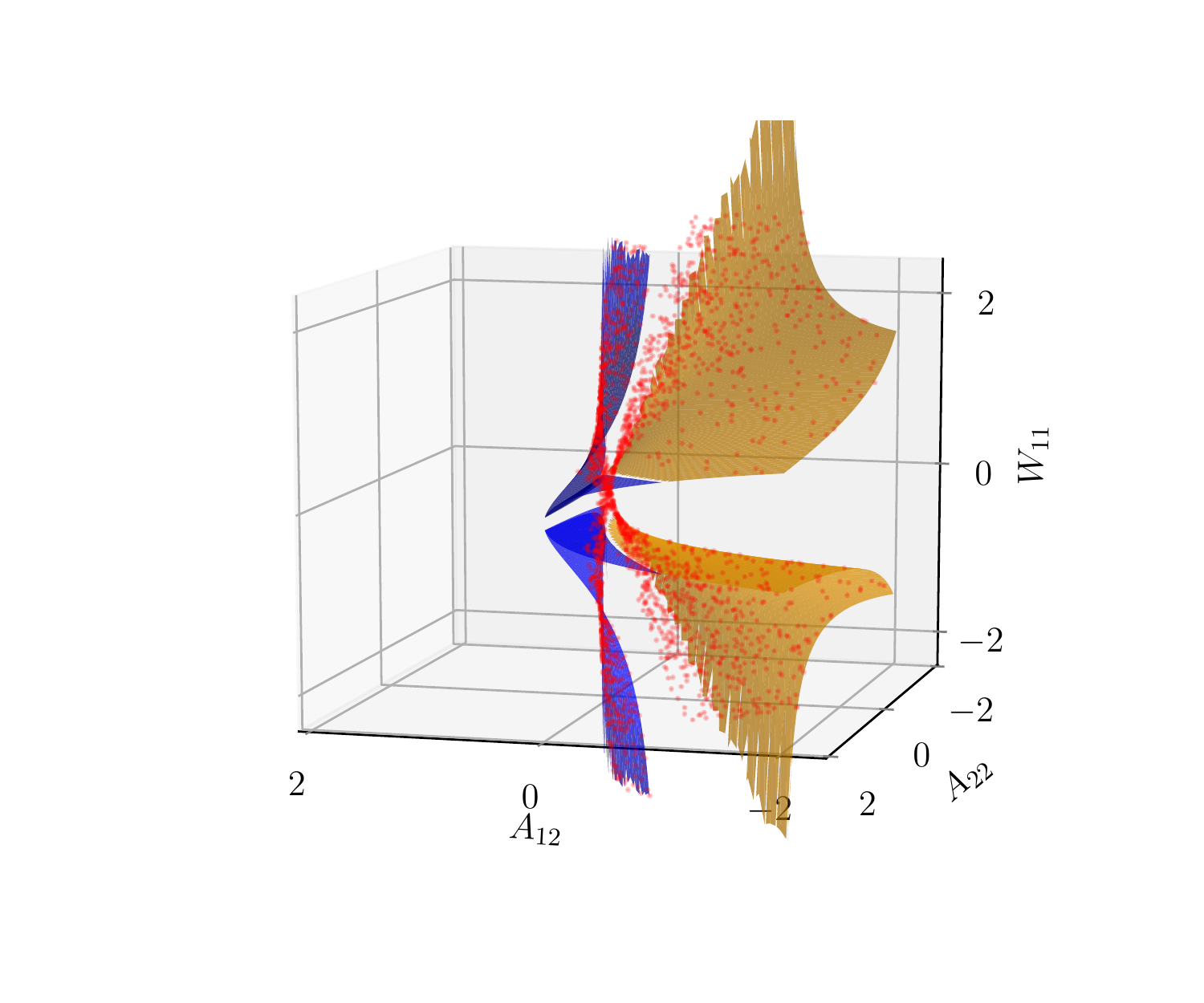}
\caption{\label{fig: 3D recurrent} Space of solutions to the recurrent (n=2) problem in the space of the two unconstrained elements of $A$: $A_{12}$ and $A_{22}$ and one element of $W$: $W_{11}$. The surfaces are analytical solutions and dots are solutions obtained numerically by gradient descent from random initial conditions.}
\end{figure}

To visualize the branching of solutions in the recurrent problem we can plot them in 3D (Figure \ref{fig: 3D recurrent}), where the added axis of $W_11$ allows us to see the multiple branches arising from square roots of $A$.

\subsubsection{Dynamics in different regions}
\begin{figure}[ht!]
\centering
\includegraphics[width=.95\linewidth]{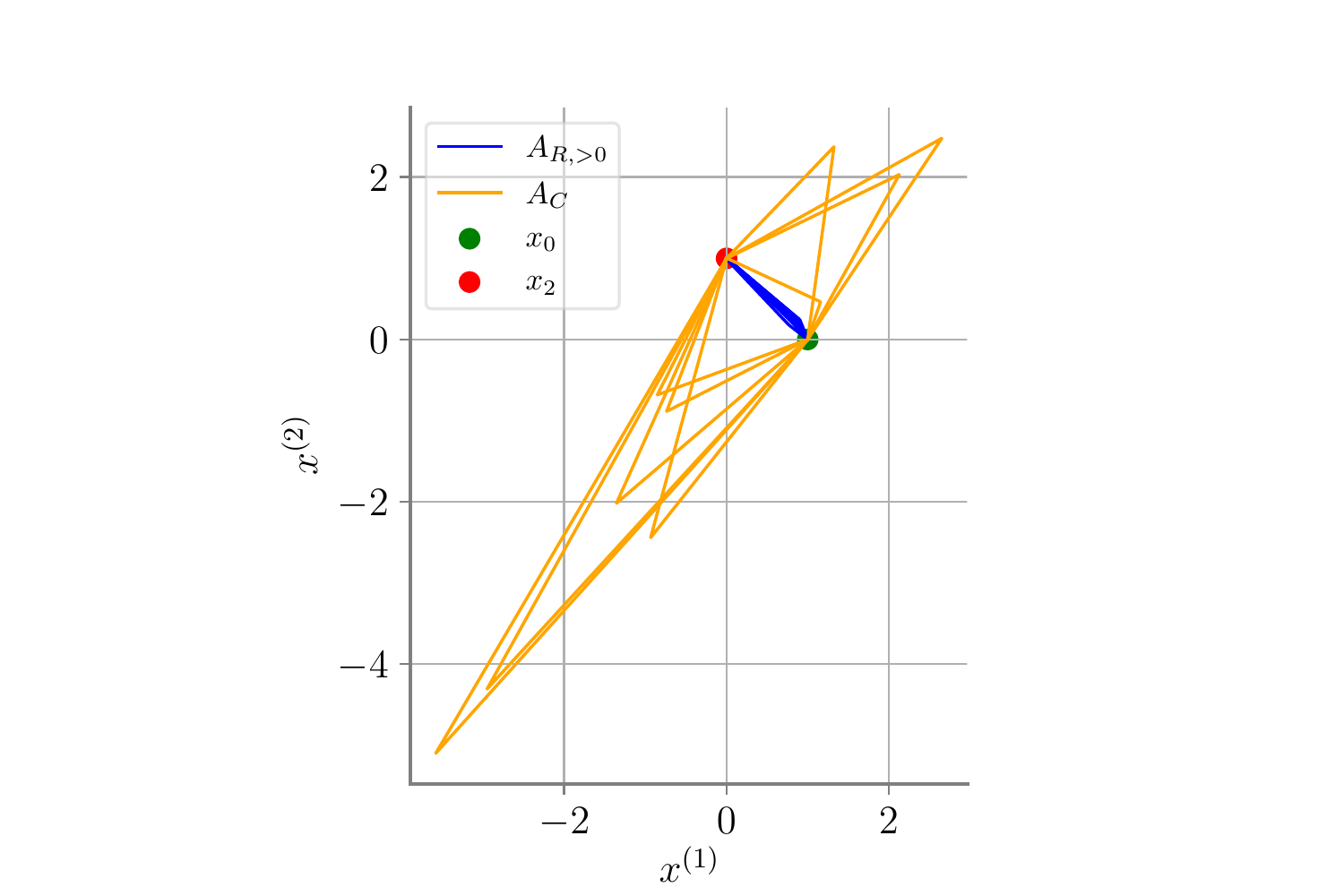}
\caption{\label{fig: dynamics}Trajectories of recurrent solutions (n=2 time steps) from the two classes $\mathcal A_{\mathbb C}$ and $\mathcal A_{\mathbb R,>0}$.}
\end{figure}

The above analysis indicated that the solution space can be divided into different regions. But are solutions from these different regions also characterized by different dynamics? Figure \ref{fig: dynamics} shows this is indeed the case, by sampling solutions from the two regions and plotting their trajectories.

\subsection{Nonlinear two-neuron RNN details and ReLU version}
\subsubsection{Training methods}
The four parameters, $W\in \mathbb{R}^{2\times 2}$, were iid sampled from a uniform distribution  $\mathcal U(-1.5,1.5)$. We implemented the continuous time dynamics of the RNN in \textit{PyTorch} \cite{Paszke2017AutomaticDI} using the package \href{https://github.com/rtqichen/torchdiffeq}{\textit{torchdiffeq}} \cite{NEURIPS2018_69386f6b,chenLearningNeuralEvent2021} enabling the calculation of gradients $\frac{\partial \mathcal L}{\partial W}$ using back-propagtion through time. We trained parameters using \textit{Adam} \cite{kingmaAdamMethodStochastic2017}.


\subsubsection{Different nonlinearity}
\begin{figure}[h]
    \centering
    \includegraphics[width=\textwidth]{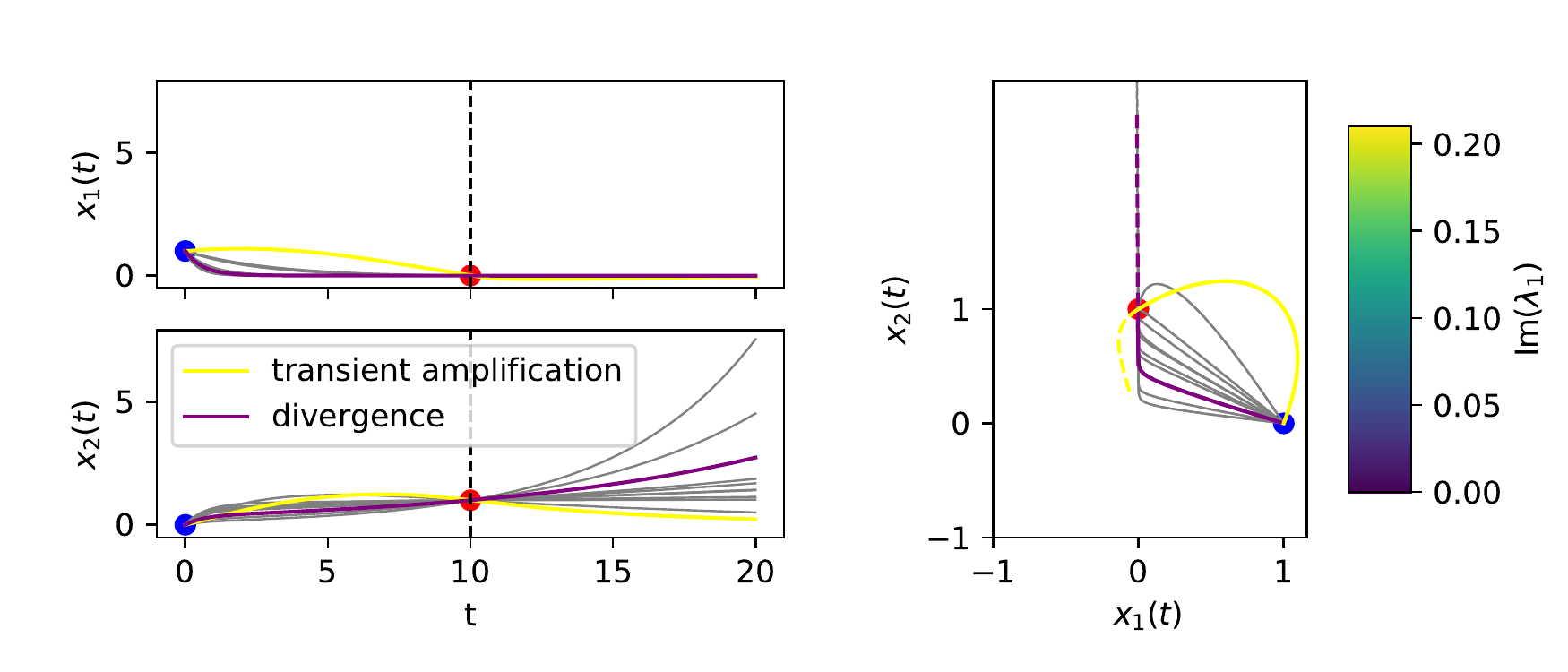}
    \caption[]{\label{fig:2DReLUtrajectories}Trajectories of several 2D RNN solutions for $\phi:=\text{ReLU}$.}
\end{figure}

\begin{figure}[h]
    \centering
    \includegraphics[width=\textwidth]{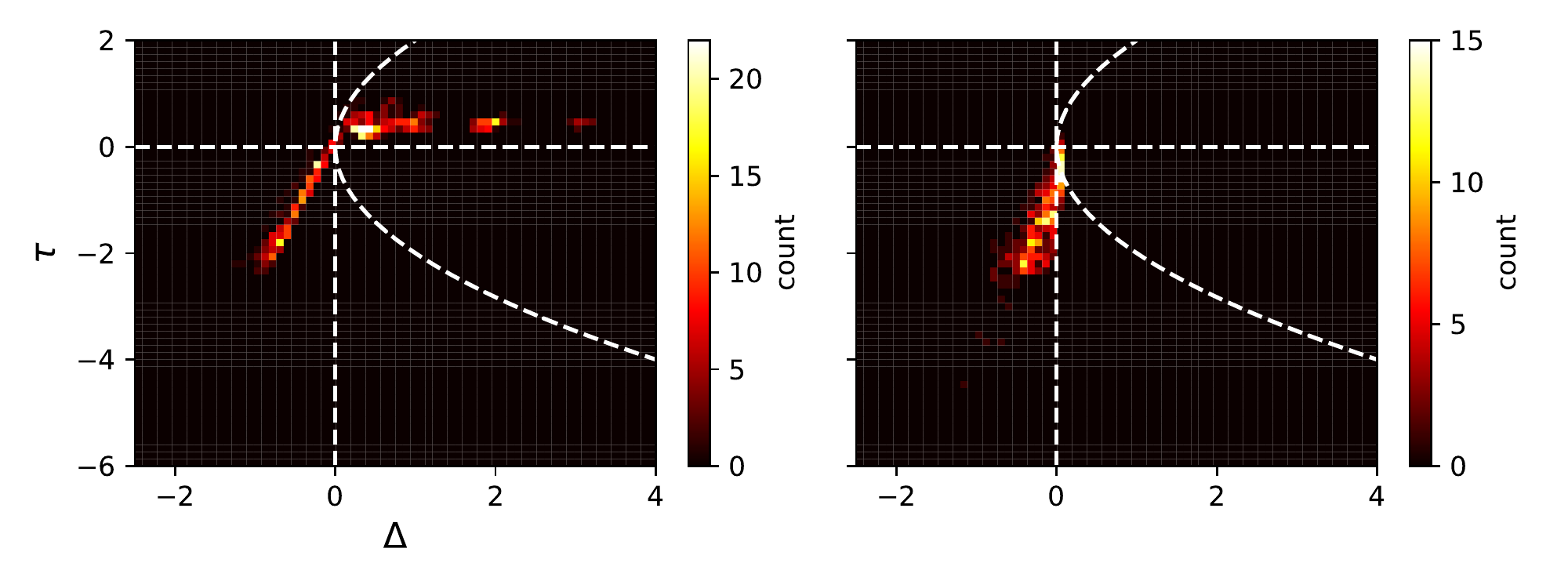}
    \caption[]{\label{fig:trace-det 2Dhist} A 2D histogram of trace $\tau$ and determinant $\Delta$ of $W$ of the 2D RNN solutions for the task specified in the main text. $\phi:=\tanh$ (left) and $\phi:=\text{ReLU}$ (right).}
\end{figure}

To examine the effect of training hyperparameters on the space of solutions, we used $\phi:=\text{ReLU}$ instead of $\phi:=\tanh$ that was used in the main text. We find that this choice indeed leads to different solution types. Specifically, ReLU RNNs did not converge to limit-cycle solutions. Some converged to non-zero fixed points, accompanied by a saddle point at the origin, as in the main text. In addition, two other solution types arose in this setting. A stable origin with large transient amplification, as in the yellow curve of Figure \ref{fig:2DReLUtrajectories}), and a diverging trajectory, shown by the dark curve in the same figure. The effect of the different non-linearity is also seen in the distribution of trace and determinants of the solutions (Figure \ref{fig:trace-det 2Dhist}), where limit cycles are absent for ReLU, and stable solutions (bottom-right quadrant) are present but uncommon for $\tanh$.

\section{Neuroscience Tasks}
\subsection{Training process}
\subsubsection{Network architecture}\label{appendix:architecture}
We studied three different RNN architectures and their exact equations are all summarized below.
The trained parameters are the weights ${W}$ and biases ${b}$. The function $\sigma(z) = (1+exp(-z))^{-1}$ is the sigmoid function,
$h_t\in\mathbb{R}^N$ and $u_t\in\{0,1\}^2$ are the state and the input at time $t$.
\paragraph{Vanilla \cite{elmanFindingStructureTime1990}}
\begin{gather}\label{eq:vanilla}
    h_t = \tanh\left(W_{ih}{u}_t + {b}_{ih}+W_{hh}h_{t-1} + b_{hh}\right)
\end{gather}
\paragraph{GRU \cite{choLearningPhraseRepresentations2014}}
\begin{align}\label{eq:gru}
    r_t&=\sigma\left(W_{ir}u_t + b_{ir} + W_{hr}h_{t-1}+b_{hr}\right)\\
    z_t&=\sigma\left(W_{iz}u_t + b_{iz} + W_{hz}h_{t-1}+b_{hz}\right)\\
    n_t&=\tanh\left(W_{i    n}u_t+b_{in}+r_t*\left(W_{hn}h_{t-1}+b_{hn}\right)\right)\\
    h_t&=(1-z_t)*n_t + z_t*h_{t-1}
\end{align}
\paragraph{LSTM \cite{hochreiterLongShortTermMemory1997}}
\begin{align}\label{eq:lstm}
    i_t&=\sigma\left(W_{ii}u_t + b_{ii} + W_{hi}h_{t-1}+b_{hi}\right)\\
    f_t&=\sigma\left(W_{if}u_t + b_{if} + W_{hf}h_{t-1}+b_{hf}\right)\\
    g_t&=\tanh\left(W_{ig}u_t+b_{ig}+r_t*\left(W_{hg}h_{t-1}+b_{hg}\right)\right)\\
    o_t&=\sigma\left(W_{io}u_t + b_{io} + W_{ho}h_{t-1}+b_{ho}\right)\\
    c_t&=f_t*c_{t-1} + i_t*g_t\\
    h_t&=o_t*\tanh(c_t)
\end{align}
The units had  $N=20,\dots,50$ hidden neurons and the output of the network at every time-step is an affine readout of the internal state. $h_0$ was always initialized to zero.

\subsubsection{Task and trial structure}\label{appendix:taskdata}
In all of the trials below, there are two input channels, one for each input pulse,  and one output channel.
Both inputs and the required output were binary sequences with ones during each pulse and zero elsewhere.
Each task is divided into seven epochs - before the first pulse, the first pulse, between pulses, the second pulse, before output pulse, output pulse, and after output pulse.
\paragraph{Delayed Discrimination  \cite{romoNeuronalCorrelatesParametric1999a}}
The first pulse with amplitude $f_1\in[2,10]$ arrives after five steps. The second input pulse with amplitude $f_1\in[2,10]$,where $f_1\neq f_2$ arrives after $5+t_d$, where $t_d\in [0,24]$.  After $15$ additional steps, the network is supposed to respond with a five-steps output pulse with amplitude $\text{sign}(f_2-f_1)$.
\paragraph{Interval Discrimination \cite{biUnderstandingComputationTime2020}}
The first pulse is given after $t_1$ steps, where $t_1\in [10,30]$. The second pulse is given $t_2$ steps after $t_1$. Both pulses last for two steps and have unit amplitude. After $15$ additional steps, the network is supposed to respond with two-steps output pulse with amplitude $\text{sign}(t_1-t_2)$.
\paragraph{Interval Reproduction \cite{jazayeriTemporalContextCalibrates2010}}
The \emph{Ready} pulse was given after $10-20$ steps. 
When working with intervals from the range $[t_{in}^{min}, t_{in}^{max}]$, the length of all trials was set to $2*t_{in}^{max}+100$. This allowed the network time to relax back to rest for at least $70$ steps after emitting a \emph{Go} pulse. All pulses were $5$ steps long. For training $t_{in}^{min}=20$ and $t_{in}^{max}=50$

The training set always included $512$ random trials so, on average, every interval was included more than $5$ times.

\subsubsection{Training protocol}\label{appendix:training} All networks were trained using \emph{Adam} \cite{kingmaAdamMethodStochastic2017} for $10000$ epochs with a batch size of $64$ and a decaying learning rate starting from $1e-3$ up until $1e-5$. Unless stated otherwise, the training set
was comprised of $512$ trials and their order was shuffled at the beginning of each epoch. We estimated the network's performance with mean squared error (MSE), and training was halted when the minimal threshold of $10^{-4}$ was achieved over the training set.

\subsection{Reduced dynamics}\label{subsec:reduceddynamics}
As we discussed in the main text, defining what is a solution is not trivial. We follow the dynamical system approach \cite{vyasComputationNeuralPopulation2020,barakRecurrentNeuralNetworks2017} and wish to obtain a compact description of the dynamics of the network and their relation to behavioral inputs and outputs. Previous work mostly focused on fixed points and their vicinity \cite{sussilloOpeningBlackBox2013,NEURIPS2019_5f5d4720}. Because transient dynamics are at the heart of some of the tasks studied here, we opted for a different approach. 
To this end, we devised a tool that, given the network weights and task inputs, builds a directed graph that captures the essence of the calculation, which we call reduced dynamics.
This process can be divided into two steps; representation of the dynamics as a directed graph, and pruning all irrelevant information from it. We will describe each of these steps next.

\subsubsection{Dynamics to Directed Graph}
Each set of dynamical trajectories and input-driven transitions between them can be interpreted as a directed graph, where each node holds a state and its corresponding readout as attributes, and each edge is weighted according to the input it represents.
However, this representation is highly redundant because autonomous trajectories converge often to a lower-dimensional dynamical object. Hence, a more exact graph representation of the dynamics would recognize trajectories at the point of convergence as identical.
Using this observation, we derived an iterative process to represent the dynamics of a network faithfully with a graph:

\begin{enumerate}
    \item Create a graph from the autonomous trajectory of the network, from a chosen initial state.
    \item Inject task inputs to states at the appropriate locations (see below) and obtain the immediate states afterward.
    \item For each such state:
\begin{enumerate}
    \item Run the dynamics from an initial state until a cycle is found or the neural speed is low.
    \item Create a subgraph from that trajectory.
    \item Connect the subgraph to the graph by the appropriate edge.
    \item Try and merge the subgraph to other previously-existing branches.
\end{enumerate}
\end{enumerate}

There is a choice to make on how to collect candidate trajectories for this process. In principle, one could follow the autonomous dynamics, and from each state inject every possible input, and follow their outcome and so forth. Such a procedure can lead to an exponential increase in the number of trajectories and is thus not feasible. We chose to extend the autonomous parts of trajectories beyond those required by the task. The inputs, however, were only provided in a task-consistent manner. For example, in the interval reproduction task, we gave only one Ready pulse and one Set pulse for each \textit{trial}.
The product of this process is a directed graph that faithfully represents the dynamics and contains the least information possible.
\paragraph{Branch-Merging} The algorithm tries to match each new branch to the previously existing ones. If the branch contains a cycle, it would be compared to all existing cyclic components. Otherwise, it will be compared to non-cyclic components. In any case, for each such component, we estimate the convergence-score of each pair of states, according to the formula:
$$
\text{score}(h^1_{t_1},h^2_{t_2})=\frac{2*\norm{h^1_{t_1}-h^2_{t_2}}_2}{q(h^1_{t_1})+q(h^2_{t_2})}, \quad h^1_{t_1},h^2_{t_2}\in\mathbb{R}^n
$$

where we quantified the neural velocity in phase space \cite{sussilloOpeningBlackBox2013} as

\begin{equation}\label{eq:q}
    q(h_t) = \Vert h_{t+1}-{h_t}\Vert_2, \quad t\in\mathbb{N}
\end{equation}

Intuitively, the score measures the separation between trajectories using their velocity as the unit of measurement. A lower score indicates a better chance for convergence.
By storing all such pairs in a matrix, we can look for sub-lower diagonals that contain values below a threshold as an appropriate candidate. In the end, the chosen branch will have the earliest connection point with the new branch. Importantly, the threshold can be tuned and used for different purposes - a lower threshold would generally lead to a higher resolution of the dynamics, and vice versa.
\subsubsection{Full dynamics graph to a reduced dynamics graph}
\begin{figure*}[h]
    \begin{center}
        \includegraphics[width=\textwidth]{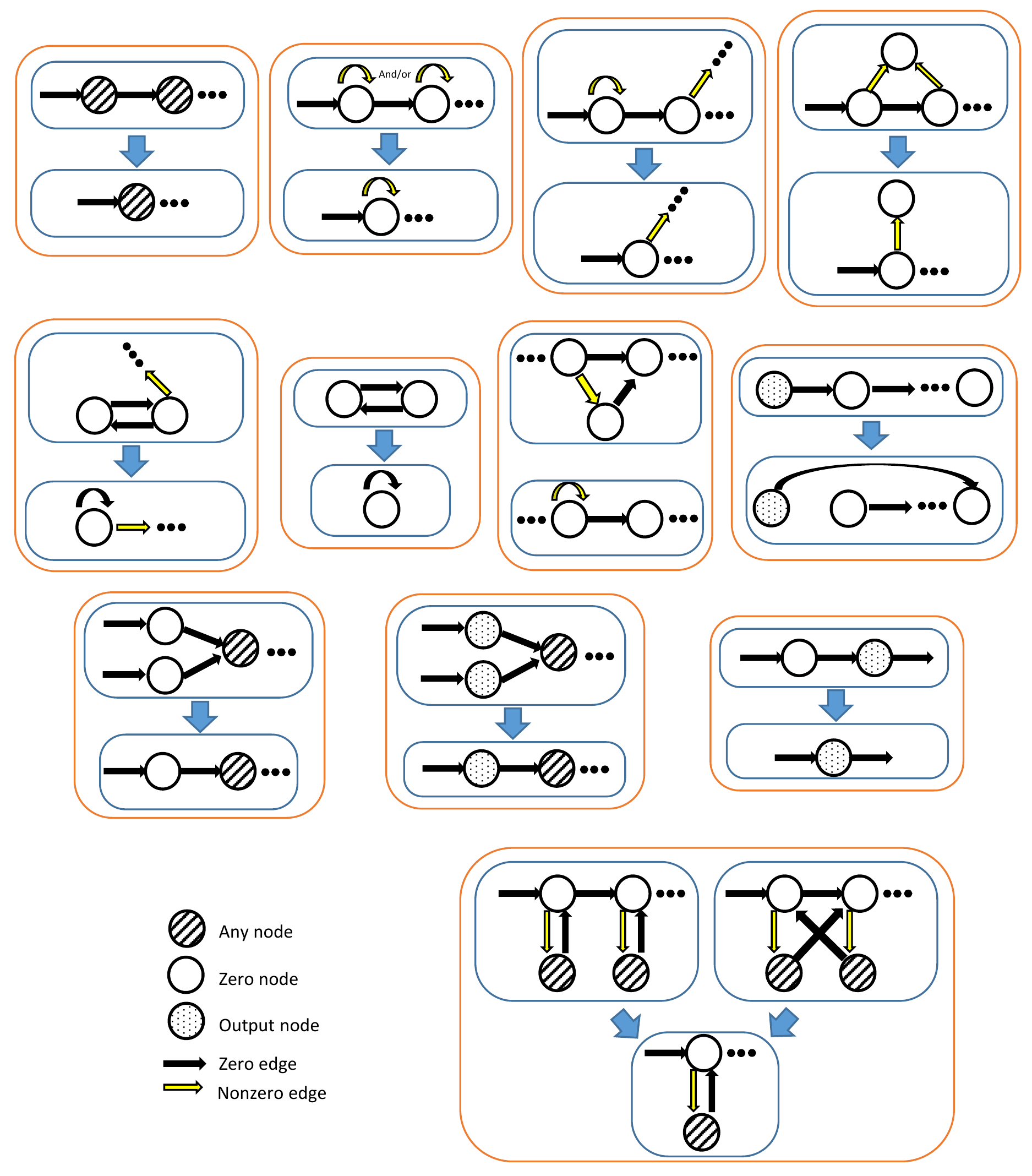}
    \end{center}
    \caption{Reduced-dynamics rules}
    \label{fig:reducedrules}
\end{figure*}

\begin{figure*}[h]
    \begin{center}
        \includegraphics[width=\textwidth]{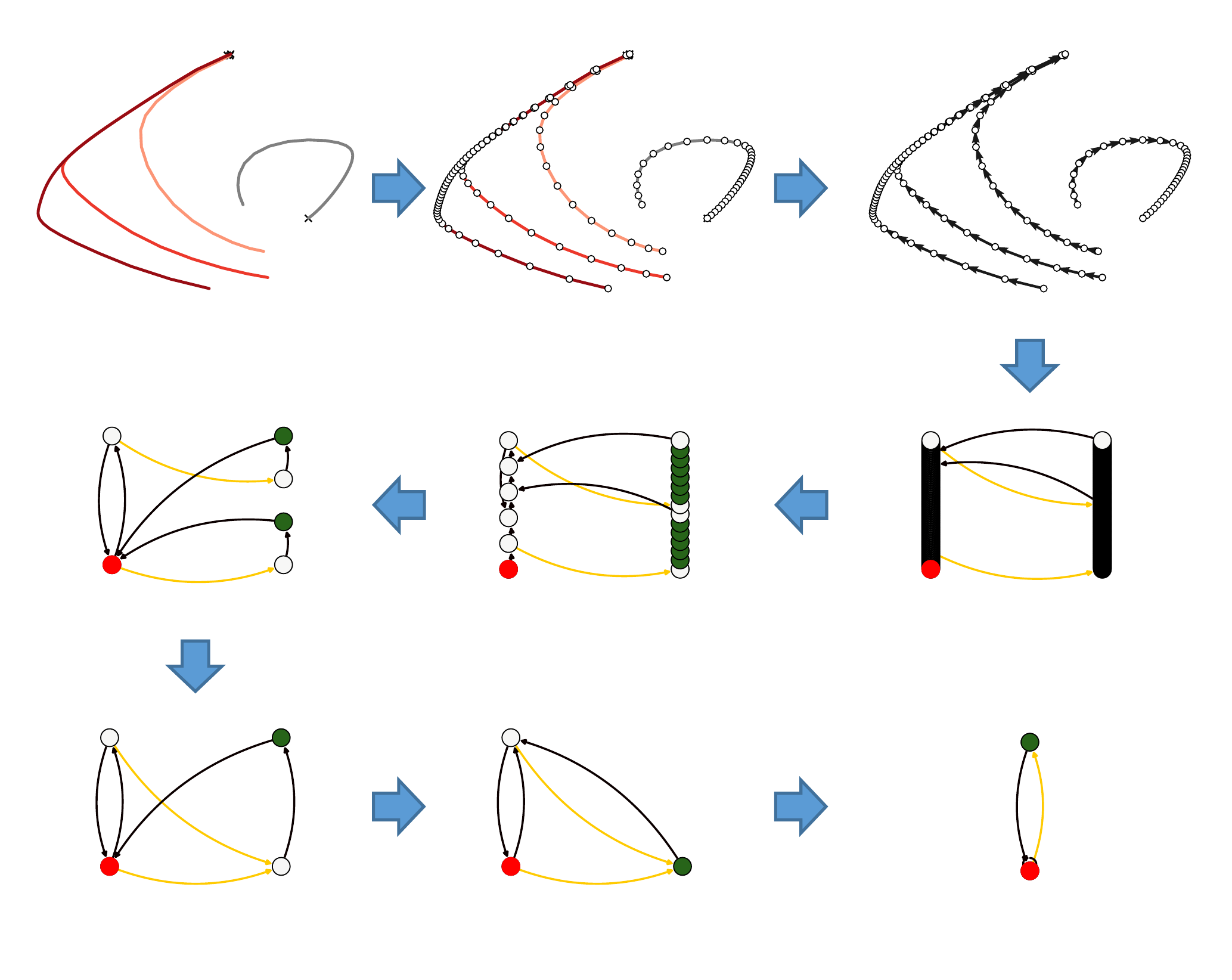}
    \end{center}
    \caption{Stages of reduced-dynamics applied to an example network}
    \label{fig:reducedexample}
\end{figure*}

In this part, we remove parts from the graph iteratively until we reach an irreducible representation that contains the essence of the computation. To achieve this, we created a list of rules that operate on a local level - structures of up to four nodes. Each rule specifies a different condition under which a structure can be compressed, without causing information loss. The main rules are shown in \cref{fig:reducedrules}, and the exact implementation is deposited in the accompanying code. Stages of the process are shown for a specific network in \cref{fig:reducedexample}. Note that this is not a \textit{lossless} compression; Some properties of the solution do not remain in the final form. For instance, the temporal distance between different states is eliminated. For the interval reproduction task, this does not allow to "see" the logic of the task from the graph, but it does allow to differentiate the major solution categories as described in the main text. The rules can be modified to include such information, and in general, the set of rules is flexible so that the user can define what type of information will remain in the final product.

\subsection{Feature extraction}\label{featureextraction}


As described in the main text, we extracted various features from the neural activity from zero-input epochs within the training set. These were related to the major dynamical objects and task epochs. For each task, epochs corresponding to one- or two- dimensional manifolds were extracted as detailed below.

\paragraph{Interval reproduction task} We extracted the 1D trajectory between \emph{Ready} and \emph{Set} (until $t_{in}^{max}$) that is shared across trials, and a 2D manifold by extracting for each task parameter $t_{in}$ its following \emph{Set}-\emph{Go} trajectory and combining the results.
\paragraph{Interval discrimination task} We extracted the 1D trajectory during the first interval (until $t_{1}^{max}$) that is shared across trials, and a 2D manifold by extracting for each task parameter $t_{1}$ the activity during the interval that follows it and combining the results.
\paragraph{Delayed discrimination} We extracted for each task parameter ($f_1$) the trajectory during the delay that follows it, and combined the results into a 2D manifold.
The two trajectories between the second pulse and the output of the network corresponding to $(10,2)$ and $(2, 10)$. In total one 2D object and two 1D objects.
We extracted features from each 1D trajectory, from each 2D manifold, and similarity measures between pairs of 1D and 2D objects.

\paragraph{1D features} The shape of a trajectory can indicate whether the network will eventually converge to a limit cycle. We thus considered the minimal and maximal curvature, the speed at its end, and the ratio between its initial and final speed. All these features were measured on a logarithmic scale. 
\paragraph{2D features}
Because this is a 2D manifold (time by trials), we calculated the aspect ratio as follows. The nominator was the cumulative length of the trajectory corresponding to the initial states across all trials. The denominator was the length of the full trajectory of the longest trial. Similarly, we extracted the aspect ratio to the final states of the manifold.
Later, we calculated how the length of each trajectory within the manifold changes as a function of the task parameter ($p$) it corresponds to, by fitting a linear regressor to the mapping $p\rightarrow ||\text{Manifold}(p)||_2^2$ and extracting its slope as a feature. For each trajectory within the manifold, we left out the first and last five time steps.
\paragraph{Cross epoch features}
Here, we focused on the relationship between the trajectory of the one-dimensional epoch and the longest trajectory out of the 2D epoch. We extracted as features the Pearson correlation and the angle between them, the ratio between their speeds, and the margin of their separating hyper-plane obtained from Linear SVM. 



\subsection{Different views of same object}\label{sec:differentviews}
\begin{figure*}[h]
    \begin{center}
        \includegraphics[width=\textwidth]{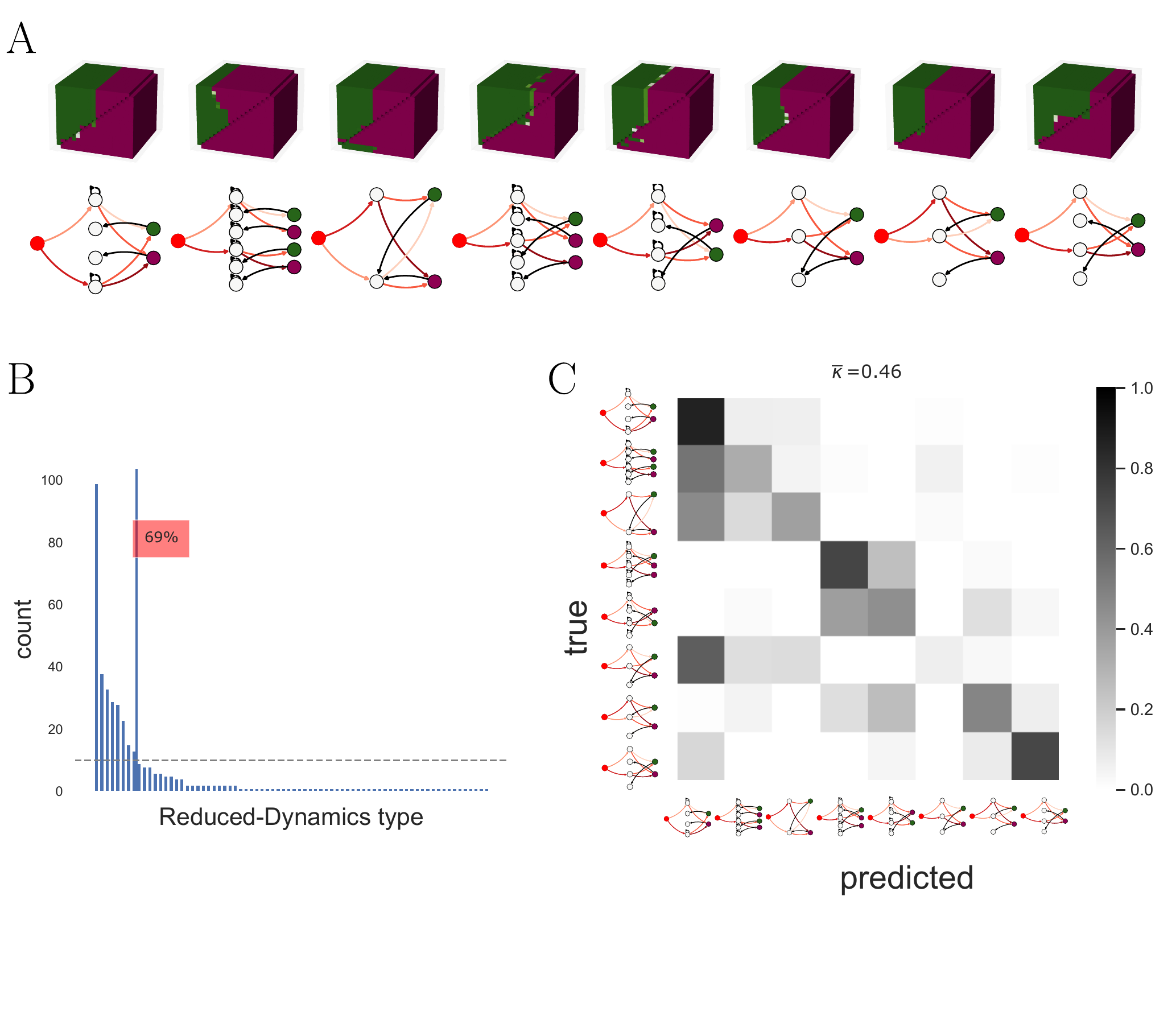}
    \end{center}
    \caption[Space of solutions]{The space of solutions for the delayed discrimination task.  \textbf{A} Representative extrapolation plots (top) and reduced dynamics graphs (bottom) for the eight most common solution types. \textbf{B} Distribution of solution types for the $400$ networks trained. The eight solutions shown account for $69\%$ of the networks. \textbf{C} Neural features obtained during the training set can partially predict the solution type that includes extrapolation dynamics. The confusion matrix shows the result of this prediction.  }
    \label{fig:romoconfusion}
\end{figure*}

\begin{figure*}[h]
    \begin{center}
        \includegraphics[width=\textwidth]{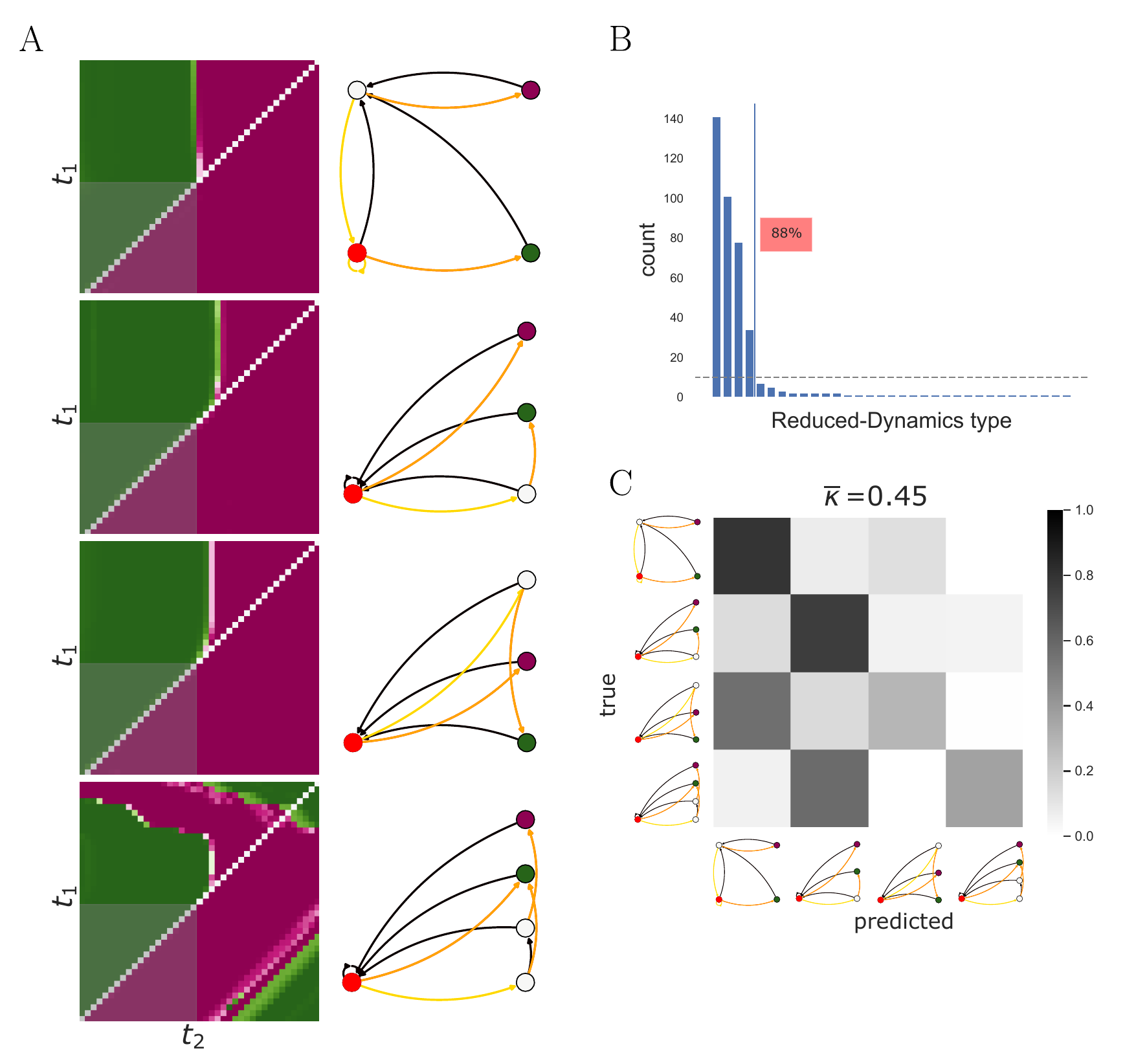}
    \end{center}
    \caption[Space of solutions]{The space of solutions for the interval discrimination task.  \textbf{A} Representative extrapolation plots (left) and reduced dynamics graphs (right) for the four most common solution types. \textbf{B} Distribution of solution types for the $400$ networks trained. The four solutions shown account for $88\%$ of the networks. \textbf{C} Neural features obtained during the training set can partially predict the solution type that includes extrapolation dynamics. According to the confusion matrix, the classifier is able to discriminate solutions that contain limit cycles from the ones which do not.   }
    \label{fig:tromoconfusion}
\end{figure*}

\begin{figure*}[h]
    \begin{center}
        \includegraphics[width=\textwidth]{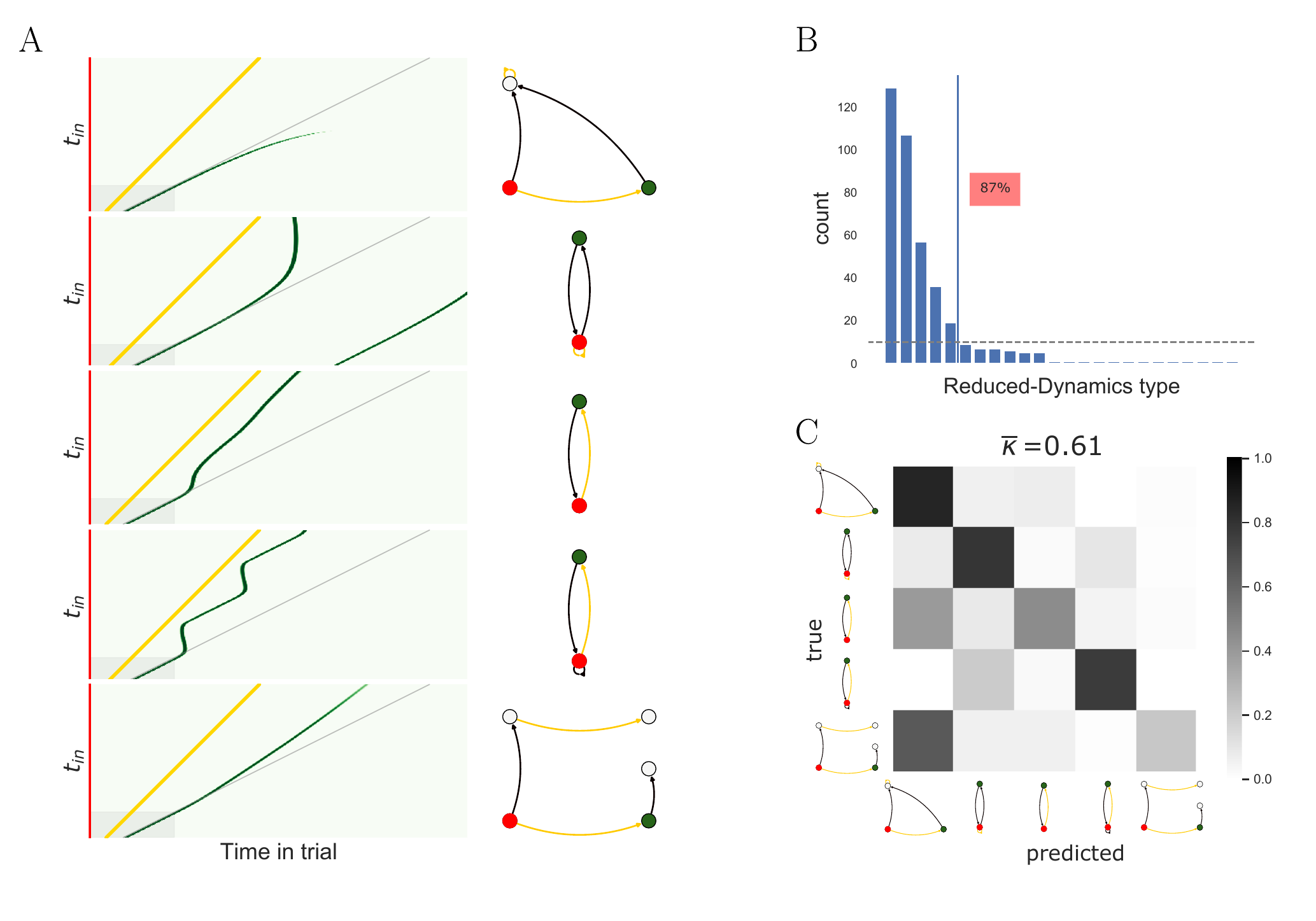}
    \end{center}
    \caption[Space of solutions]{The space of solutions for the time reproduction task, for GRU networks.  \textbf{A} Representative extrapolation plots (left) and reduced dynamics graphs (right) for the five most common solution types. \textbf{B} Distribution of solution types for the $400$ networks trained. The five solutions shown account for $87\%$ of the networks. \textbf{C} Neural features obtained during the training set can partially predict the solution type that includes extrapolation dynamics. The confusion matrix shows the result of this prediction. }
    \label{fig:gruconfusion}
\end{figure*}

\begin{figure*}[h]
    \begin{center}
        \includegraphics[width=\textwidth]{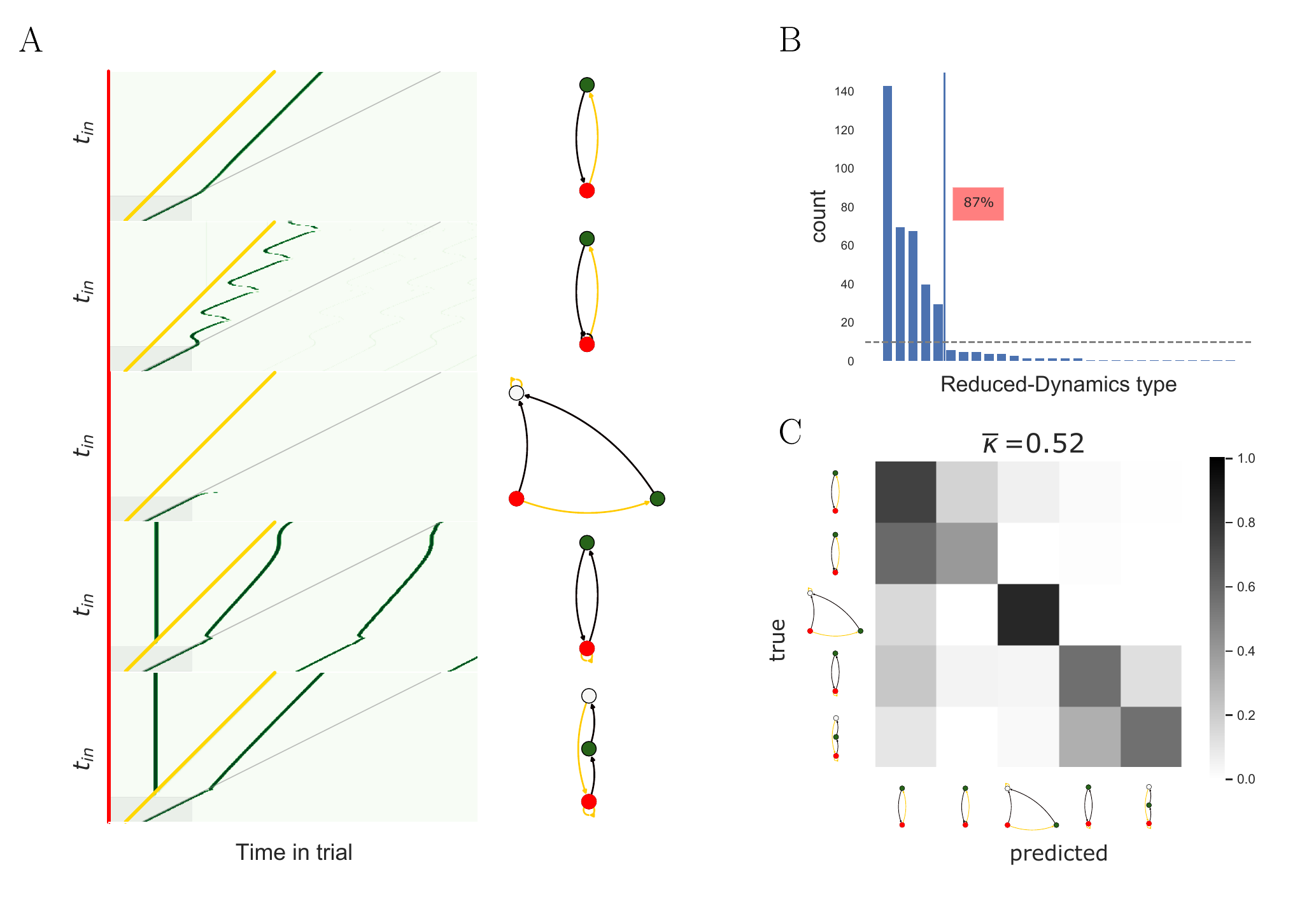}
    \end{center}
    \caption[Space of solutions]{The space of solutions for the time reproduction task, for LSTM networks.  \textbf{A} Representative extrapolation plots (left) and reduced dynamics graphs (right) for the five most common solution types. \textbf{B} Distribution of solution types for the $400$ networks trained. The five solutions shown account for $87\%$ of the networks. \textbf{C} Neural features obtained during the training set can partially predict the solution type that includes extrapolation dynamics. The confusion matrix shows the result of this prediction. }
    \label{fig:lstmconfusion}
\end{figure*}

To see whether the neural data from the training set contains information about the topology of the networks, we evaluated the ability of the neural features to predict the reduced dynamics we described earlier.
This was done by a cross-validation procedure that included $50$ repetitions of fitting a Random-Forest classifier to a randomly selected $90\%$ training set of the networks. We evaluate the \emph{Kappa-Cohen} score and the confusion matrix of the classification on the remaining $10\%$. The results of averaging across all repetitions appear on \cref{fig:romoconfusion,fig:tromoconfusion,fig:gruconfusion,fig:lstmconfusion} and in the main text.

\section{Variability in Context-dependent integration}

\begin{figure}[h]
    \begin{center}
        \includegraphics[width=0.98\textwidth]{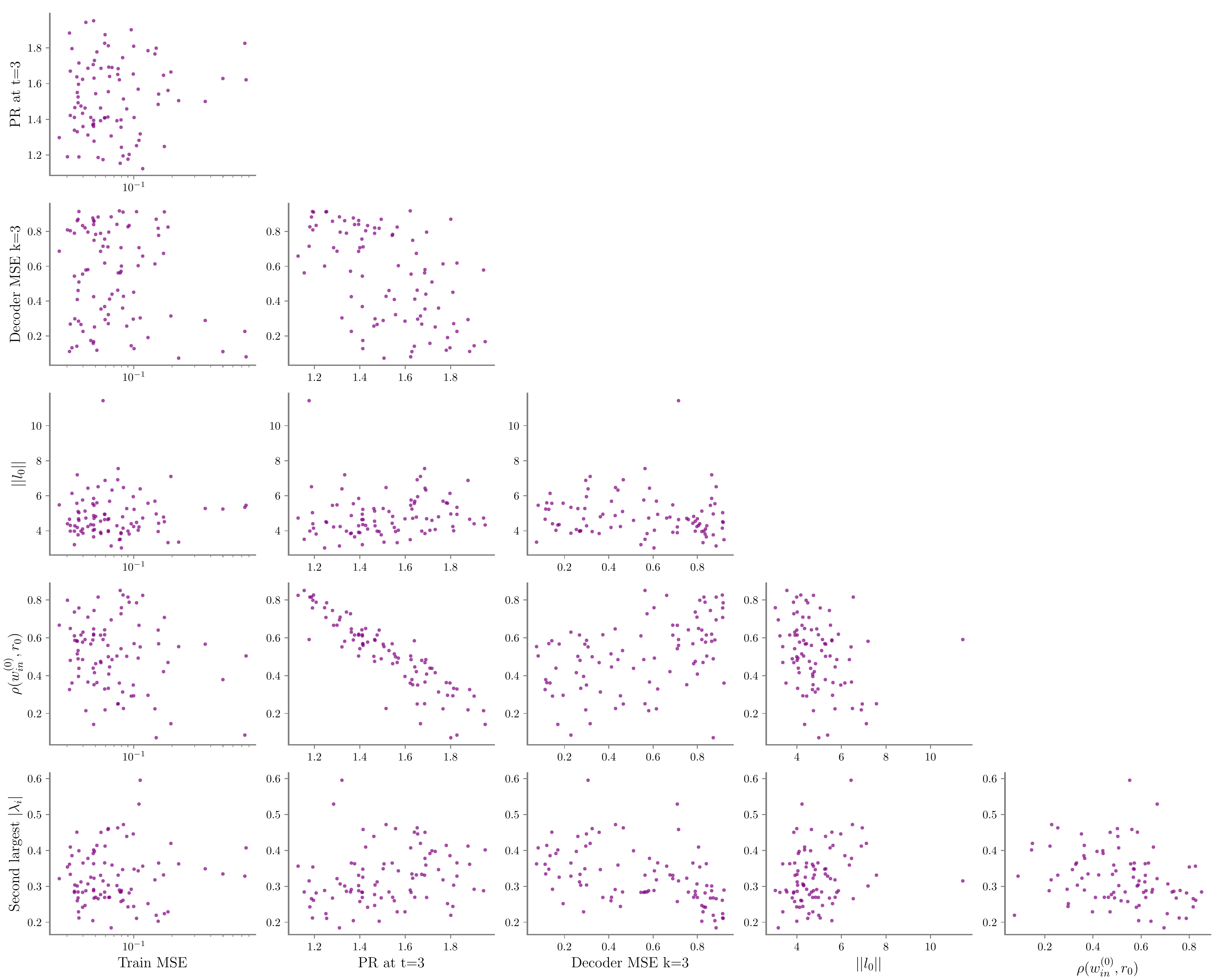}
    \end{center}
    \caption{Comparison of neural measurements along with training MSE loss against each other across independently initialized Vanilla networks (25,30,35 hidden units) networks trained on context dependent integration (CDI) task.\cite{manteContextdependentComputationRecurrent2013,NEURIPS2019_5f5d4720} Each dot corresponds to measurements on a single network.}
    \label{fig:CDIneuralmeasures}
\end{figure}

Not all tasks display qualitative variability \cite{NEURIPS2019_5f5d4720}.
Even without such variability, there can be substantial quantitative variability. Here we highlight several features of such variability (\cref{fig:CDIneuralmeasures}).

We train several independently initialized vanilla \eqref{eq:vanilla} networks on the context-dependent integration (CDI) task, following the task protocol from  \cite{NEURIPS2019_5f5d4720}. All networks form an approximate line attractor  \cite{NEURIPS2019_5f5d4720}. 
For each network, we inject zero channel input along with one of the context, and thus allow the network to converge to the origin of the line attractor \cite{sussilloOpeningBlackBox2013}. 
Following previous works, we analyze the linearized system around this point. We characterize the network with the following neural measurements.
\begin{enumerate}
    \item Participation ratio (PR) at t=3: to the linearized system, we deliver an input $u_t\sim \mathcal N(0,1)$ for t consecutive time steps and compute the participation-ratio, a measure of linear dimensionality defined as $\frac{(\sum_i \lambda_i)^2}{\sum_i \lambda_i^2}$ where the $\lambda_i$s are the eigenvalues of $C_t=<h_th_t^T>_{\text{trials}}$ of the network hidden-state at time $t$ across several trials.
    \item Decoder MSE k=3: to the linearized system, we deliver $u_t\sim \mathcal N(0,1)$ for $T=30$ time steps and we perform linear regression to decode $u_{T-k}$ from $h_T$. We use the decoder MSE as a proxy measure for information held by the network about previous inputs.
    \item $\norm{l_0}$: norm of the left eigenvector of the linearized system corresponding to eigenvalue with the largest absolute value, i.e the selection vector \cite{manteContextdependentComputationRecurrent2013}.
    \item $\rho(w_{in}^{(0)},r_0)$: Correlation of input weight vector $w_{in}$ with the right eigenvector of the linearized system corresponding to eigenvalue with the largest absolute value, i.e. the direction along the line attractor \cite{manteContextdependentComputationRecurrent2013}.
    \item Second largest $|\lambda_i|$: Eigenvalue of linearized recurrent dynamics with second largest absolute value.
\end{enumerate}
\cref{fig:CDIneuralmeasures} shows each of these neural measures compared to each other.

\begin{figure}[h]
    \begin{center}
        \includegraphics[width=0.98\textwidth]{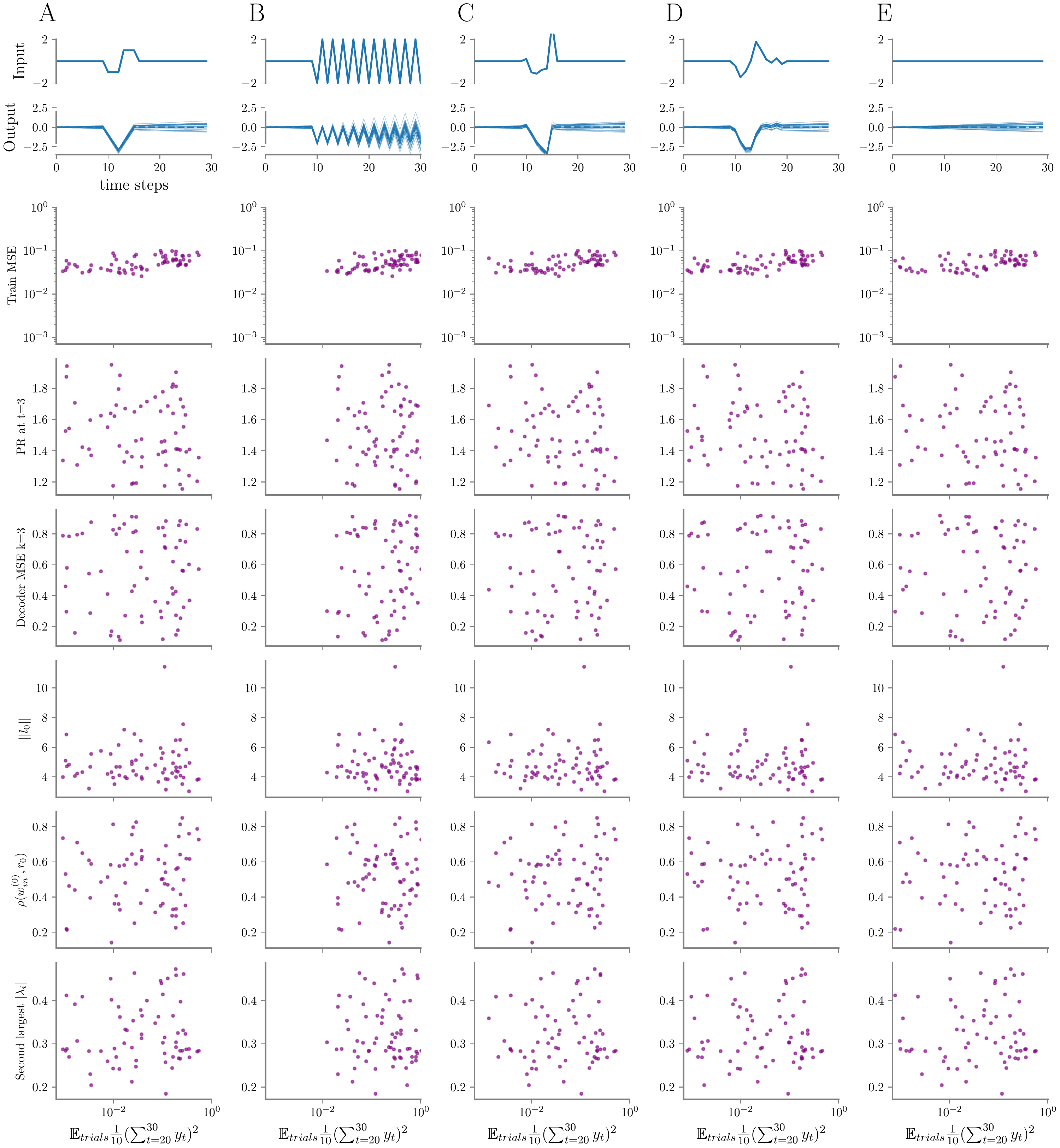}
    \end{center}
    \caption{Comparison of neural and behavioral measurements for RNNs trained on CDI (see \cref{fig:CDIneuralmeasures} for details about the colors). We run the networks on five different input protocols, one in each of the columns A,B,C,D,E. First two rows show example inputs and outputs/targets for each of the tasks. Rows 3 to 8 show behavioral MSE for each task versus the neural measurements from \cref{fig:CDIneuralmeasures} wherein each dot indicates measurements for a single network.}
    \label{fig:CDIneuralbehavioral}
\end{figure}

In addition we evaluate several behavioral measures. We run the networks on a selection of designed inputs. In all cases, we provide input only along one of the channels along with the corresponding context input. The five choices of channel inputs are shown in the top row of \cref{fig:CDIneuralbehavioral} (A,B,C,D,E). The inputs are designed such that integral is zero at the end, to facilitate visualization of the error In each case we evaluate MSE of the output from the target of zero. We then compare this MSE with training MSE and the aforementioned neural measurements for each of the networks.

Our neural and behavioral measures indicate a large degree of quantitative variability across networks. The measurements are only weakly correlated to each other, showing that there are many axes of variability. We note that even for networks with very low training MSE, there is still considerable variability in the behavioral challenges (the horizontal nature of the second row of \cref{fig:CDIneuralbehavioral}). We note, however, that extensive training results in step-like decreases in the training MSE. These steps are accompanied by a reduction in the variability of the behavioral challenges, but the latter variability remains much larger for all cases.

\end{document}